\pdfoutput=1
\documentclass[prb,aps,twocolumn,superscriptaddress,floatfix,citeautoscript,eqsecnum]{revtex4}
\usepackage{graphicx,rotating,subfigure,amsmath,amsfonts,amssymb,delarray}
\usepackage[T1]{fontenc}
\renewcommand{\vec}[1]{\boldsymbol #1}
\newcommand{\e}{\text{e}}
\newcommand{\im}{\text{i}}
\def\l{\left}
\def\r{\right}
\def\12{\frac{1}{2}}
\def\nn{\nonumber}
\newcommand{\be}{\begin{equation}}
\newcommand{\ee}{\end{equation}}
\newcommand{\bea}{\begin{eqnarray}}
\newcommand{\eea}{\end{eqnarray}}
\newcommand{\ket}{\rangle}
\newcommand{\bra}{\langle}
\newcommand{\J}{\mathcal{J}}

\begin{document}
\bibliographystyle{apsrev}

\title{Conservation laws, integrability and transport in one-dimensional quantum systems}

\author{J. Sirker} 
\affiliation{Department of Physics and Research Center OPTIMAS, University of Kaiserslautern, D-67663 Kaiserslautern, Germany}

\author{R. G. Pereira}

\affiliation{Instituto de F\'{\i i}sica de S\~ao Carlos, Universidade de S\~ao Paulo, C.P. 369, 
S\~ao Carlos, SP.  13566-970, Brazil}

\author{I. Affleck}
\affiliation{Department of Physics and Astronomy, University of British
  Columbia, Vancouver, BC, Canada V6T1Z1}

\date{\today}

\begin{abstract}
  In integrable one-dimensional quantum systems an infinite set of
  {\it local} conserved quantities exists which can prevent a current
  from decaying completely. For cases like the spin current in the
  $XXZ$ model at zero magnetic field or the charge current in the
  attractive Hubbard model at half filling, however, the current
  operator does not have overlap with any of the local conserved
  quantities. We show that in these situations transport at finite
  temperatures is dominated by a diffusive contribution with the Drude
  weight being either small or even zero. For the $XXZ$ model we
  discuss in detail the relation between our results, the
  phenomenological theory of spin diffusion, and measurements of the
  spin-lattice relaxation rate in spin chain compounds. Furthermore,
  we study the Haldane-Shastry model where the current operator is
  also orthogonal to the set of conserved quantities associated with
  integrability but becomes itself conserved in the thermodynamic
  limit.
\end{abstract}


\maketitle

\section{Introduction}
In classical dynamics the KAM theorem quantitatively explains what
level of perturbation can be exerted on an integrable system so that
quasi-periodic motion survives.\cite{Arnoldbook} A classical system
with Hamiltonian $H$ and phase space dimension $2N$ is integrable if
$N$ constants of motion $Q_k$ exist (i.e., the Poisson bracket
vanishes, $\{H,Q_k\}=0$) which are pairwise different, $\{Q_k,Q_l\}=0$.
Defining integrability for a quantum system is, however, much
more complicated and no analogue of the KAM theorem is known. For {\it
  any} quantum system in the thermodynamic limit an infinite set of
operators exist which commute with the Hamilton operator.
This can be seen by considering, for example, the projection operators
onto the eigenstates of the system, $[H,|n\rangle\langle n|] =0$ with
$H|n\rangle =E_n|n\rangle$. In quantum systems described by
tight-binding models with short-range interactions it is therefore
important to distinguish between {\it local} conserved quantities
$Q_n=\sum_jq_{n,j}$, where $q_{n,j}$ is a density operator acting on
$n$ adjacent sites $j$, and {\it nonlocal} conserved quantities like
the projection operators mentioned above. In a field theory, a
conserved operator is local if it can be written as integral of a
fully local operator. Most commonly, quantum systems are called
integrable if an infinite set of {\it local} conserved quantities
exists which are pairwise different.  This definition includes, in
particular, all Bethe ansatz integrable one-dimensional quantum
systems. Here the local conserved quantities can be explicitly
obtained by taking consecutive derivatives of the logarithm of the
appropriate quantum transfer matrix with respect to the spectral
parameter.\cite{tak99}

In recent years, many studies have been devoted to the question if
integrability can stop a system from
thermalizing\cite{KinoshitaWenger,HofferberthLesanovsky,RigolDunjko}
or a current from decaying
completely.\cite{CastellaZotos,ZotosPrelovsek,Zotos,KluemperJPSJ,RoschAndrei,AlvarezGros,AlvarezGros2,FujimotoKawakami,JungRoschDrude,NarozhnyMillis,HeidrichMeisner2,JungHelmes,KluemperSakai,SirkerPereira}
That conservation laws and the transport properties of the considered
system are intimately connected is obvious in linear response theory
(Kubo formula), which relates the optical conductivity to the retarded
equilibrium current-current Green's function. Specializing to a
lattice model with nearest-neighbor hopping, the Kubo formula
reads\footnote{This formula also applies for a lattice model with
  longer range hopping if we replace the kinetic energy operator by
  the operator obtained by taking the second derivative of the
  Hamiltonian with respect to a magnetic flux penetrating the ring (see also
  Sec.~\ref{secHS}). } \be
\sigma(\omega)=\frac{i}{\omega}\l[\frac{\bra E_{\rm kin}\ket}{L}+\bra
\mathcal{J},\mathcal{J}\ket_{\rm ret}(\omega)\r].
\label{Kubo} 
\ee Here, $L$ is the system size, $T$ the temperature, $E_{\rm kin}$
the kinetic energy operator, $\mathcal{J}$ the spatial integral of the
current density operator, and the brackets denote thermal average. The
real part of the optical conductivity can be written as \be
\sigma'(\omega)=2\pi D\delta(\omega)+\sigma_{reg}(\omega).
\label{Drude}
\ee A nonzero {\it Drude weight} $D$ implies an infinite dc
conductivity.  Castella {\it et al.}~[\onlinecite{CastellaZotos}]
showed that it is possible to circumvent the direct calculation of the
current-current correlation function in the Kubo formula and compute
the finite temperature Drude weight using a generalization of Kohn's
formula.\cite{Kohn} This formula relates the Drude weight to the
curvature of the energy levels with respect to magnetic flux. The
observation that integrable and nonintegrable models obey different
level statistics,\cite{BerryTabor,BohigasGiannoni,PoilblancZiman} as
well as the calculation of the Drude weight from exact diagonalization
for finite size systems,\cite{Zotos1996} led to the conjecture that
anomalous transport, in the form of a finite $D(T>0)$, is a generic
property of integrable models.\cite{CastellaZotos}

This conjecture is corroborated by the relation between the Drude
weight and the long-time asymptotic behavior of the current-current
correlation function\cite{ZotosPrelovsek}
\begin{equation}
\label{Mazur}
D = \frac{1}{2LT}\lim_{t\to\infty} \langle \J(t)\J(0)\rangle
\geq\frac{1}{2LT}\sum_k\frac{\langle 
\J Q_k\rangle^2}{\langle Q_k^2\rangle} \, .
\end{equation}
In view of this relation, {\it ballistic transport}, $D\neq 0$, means
that the current-current correlation function does not completely
decay in time. The $Q_k$ operators in Eq.~(\ref{Mazur}) form
a set of commuting conserved quantities which are orthogonal in the
sense that $\langle
Q_kQ_l\rangle=\langle Q_k^2\rangle\delta_{kl}$.
That conserved quantities provide a lower bound for the long-time
asymptotic value of correlation functions is a general result due to
Mazur\cite{mazur}. In fact, it can be shown that the equality holds if
the right-hand side of Eq.~(\ref{Mazur}) includes {\it all} conserved
quantities $Q_k$, local and non-local.\cite{Suzuki}

The implications of Mazur's inequality for transport in integrable
quantum systems were pointed out by Zotos {\it et
  al.}~[\onlinecite{ZotosPrelovsek}]. For integrable models where at
least one conserved quantity $Q_n$ has nonzero overlap with the
current operator, $\langle Q_n \mathcal{J}\rangle\neq 0$, Mazur's
inequality implies that the Drude weight is finite at finite
temperatures. This happens, for instance, for charge transport in the
Hubbard model away from half-filling and for spin transport in the
$S=1/2$ $XXZ$ model at finite magnetic field. By contrast, for
nonintegrable models, for which all local nontrivial conservation laws
are expected to be broken, the right-hand side of Eq.~(\ref{Mazur}) is
expected to vanish so that $D(T>0)=0$. In these cases the delta
function in Eq.~(\ref{Drude}) is generically presumed to be broadened
into a Lorentzian Drude peak. The dc conductivity is finite and the
system is said to exhibit {\it diffusive transport}.

However, in many cases of interest the known local conserved
quantities associated with integrability are orthogonal to the current
operator because of their symmetry properties. This happens, for
instance in the $XXZ$ model at zero magnetic field $h$. In terms of
spin-1/2 operators $\mathbf{S}_l$ the model reads \be
H=\sum_{l=1}^N[J(S_l^xS_{l+1}^x+S_l^yS_{l+1}^y+\Delta
S_l^zS_{l+1}^z)-hS_l^z].
\label{HXXZspin} 
\ee Here $N$ is the number of sites, $\Delta$ parametrizes an exchange
anisotropy, and $J$ is the exchange constant which we set equal to $1$
in the following. One can show that all local conserved quantities of
the $XXZ$ model are even under the transformation $S_j^z\to-S_j^z$,
$S_j^\pm\to S_j^\mp$, whereas the current operator is
odd.\cite{ZotosPrelovsek} In these cases the integrability-transport
connection has remained a conjecture.  Nonetheless, several works have
presented support for a finite Drude weight in models where the
current operator has no overlap with the local conserved quantities,
at least in some parameter regimes. The list of methods employed
include exact diagonalization (ED),\cite{NarozhnyMillis,
  HeidrichMeisner2, JungRoschDrude, RigolShastry} Quantum Monte Carlo
(QMC)\cite{KirchnerEvertz, AlvarezGros2,
  HeidarianSorella,GrossjohannBrenig} and Bethe ansatz
(BA).\cite{FujimotoHubbard, Zotos,KluemperJPSJ}

In particular for the $XXZ$ model, ED results for chains of lengths up
to $L=18$ sites\cite{HeidrichMeisner2} suggest that at high
temperatures the Drude weight extrapolates to a finite value in the
thermodynamic limit for values of exchange anisotropy in the critical
regime, including the Heisenberg point. The Drude weight appears to
vanish for large values of anisotropy in the gapped regime, but the
minimum value of anisotropy for which it vanishes cannot be determined
precisely. Although no conclusions can be drawn about the low
temperature regime, the claim is that if the Drude weight is finite at
high temperatures it must also be finite and presumably even larger at
low temperatures.  The weakness of this method is that it assumes that
the finite size scaling of $D(T,L)$, which is not known analytically
and is only obtained numerically for $L\leq18$, can be extrapolated to
the thermodynamic limit. This is not necessarily true for strongly
interacting models where even at high
temperatures  there may be a large length scale  above which the behavior of dynamical properties changes
qualitatively.

In the QMC approach for the $XXZ$ model in
Refs.~[\onlinecite{AlvarezGros,AlvarezGros2}], the Drude weight was
obtained by analytic continuation of the conductivity function
$\sigma(q,i\omega_n)$, which is a function of the Matsubara
frequencies $\omega_n$, to real frequencies. This method uses a
fitting function to try to extract a decay rate $\gamma$ which
broadens the Drude peak if the Drude weight vanishes, but it clearly
fails to find decay rates that are smaller than the separation between
Matsubara frequencies, $\gamma\ll T$. In fact, the application of this
method has even led to the conclusion that the Drude weight is finite
for some gapless systems that are not integrable.\cite{KirchnerEvertz,
  HeidarianSorella} This is hard to believe, considering that
Eq.~(\ref{Mazur}) requires the existence of nontrivial conservation
laws which have a finite overlap with the current operator in the
thermodynamic limit in order for the Drude weight to be finite. An
attempt was made to explain the Drude weight for nonintegrable models
described by a Luttinger liquid fixed point based on conformal field
theory\cite{FujimotoKawakami}, but this analysis neglects irrelevant
interactions that lead to current decay and render the conductivity
finite.\cite{RoschAndrei}

Although the Drude weight for the $XXZ$ model has been calculated
exactly by BA at $T=0$ using Kohn's formula \cite{ShastrySutherland},
the calculation of $D(T>0)$ by BA is hindered by the need to resort to
approximations in the treatment of the excited states. The BA
calculation by Zotos\cite{Zotos} follows an ansatz proposed by
Fujimoto and Kawakami\cite{FujimotoHubbard} that employs the thermodynamic
Bethe ansatz (TBA), which relies on the string hypothesis
for bound states of magnons.\cite{tak99} This approach predicts that
the Drude weight is finite and decreases monotonically with $T$ for
the $XXZ$ model in the critical regime at zero magnetic field, except
at the Heisenberg point, where $D(T)$ vanishes for all finite
temperatures.  Benz {\it et al.}~[\onlinecite{KluemperJPSJ}]
criticized the TBA result and pointed out that it violates exact
relations for $D(T)$ at high temperatures. These authors presented an
alternative BA calculation of the Drude weight based on the spinon and
anti-spinon particle basis and predicted a different temperature
dependence than the TBA result.  In particular, the Drude weight is
found to be finite for the Heisenberg model at zero magnetic field.
Actually, for values of anisotropy near the isotropic point this
approach predicts that $D(T)$ increases with $T$ at low temperatures.
Like the TBA result, the result based on spinons and anti-spinons
violates exact relations at high temperatures. A consistent
calculation of the Drude weight by applying the BA to the finite
temperature Kohn formula is therefore still an unresolved issue.

Integer-spin Heisenberg lattice models are not integrable, but their
low energy properties are often studied in the framework of the
continuum O(3) nonlinear sigma model, which is integrable.
Interestingly, BA calculations for the nonlinear sigma model predict a
finite $D(T>0)$ that is exponentially small at temperatures below the
energy gap.\cite{FujimotoJPSJ,Konik} In fact, it has been
argued\cite{Konik} that the finite Drude weight is due to at least one
\emph{nonlocal} conserved quantity of the quantum model which is known
explicitly\cite{Luscher} and has overlap with the current operator.
This result is not consistent with the semiclassical results by Damle
and Sachdev\cite{DamleSachdev, DamleSachdev2}, which predict diffusive
behavior for gapped spin chains at low temperatures. No nonlocal
conserved quantities that overlap with the current operator are known
explicitly for the integrable $S=1/2$ $XXZ$ model. For finite chains
such quantities can be constructed explicitly, however, in a numerical
study of chains with $L\leq 18$ no definite conclusions could be made
whether or not the overlap remains finite in the thermodynamic limit.\cite{JungRoschDrude}

The debate about the role of integrability in transport properties of
integrable models is connected with the question of \emph{diffusion}
in $S=1/2$ spin chains. The term spin diffusion first appeared in the
context of the phenomenological
theory,\cite{Bloembergen,deGennes,SteinerVillain} where it refers to a
characteristic form of long-time decay of the spin correlation
function. In the phenomenological theory, the case where the total
magnetization in the direction of quantization, $S^z=\sum_j S^z_j$, is
conserved is considered. It is then said that spin diffusion occurs if
the Fourier transform of the spin-spin correlation function
$G(\mathbf{q},t)=\sum_{j}e^{-i\mathbf{q}\cdot(\mathbf{r}_j-\mathbf{r}_0)}\langle
S^z_j(t)S^z_{0}(0)\rangle$ for small wavevector $\mathbf{q}$ decays
with time as $G(\mathbf{q},t)\sim e^{-D\mathbf q^2t}$, where $D$ is
the diffusion constant. Provided that the behavior at small $q$
dominates, this implies that the Fourier transform decays as
$G(\mathbf{r},t)
\sim t^{-d/2}\exp(-|\mathbf{r}|^2/4Dt)$.
This theory was formulated to explain results of inelastic neutron
scattering experiments in three-dimensional ferromagnets at high
temperatures. The assumptions of the theory are usually motivated by
the picture that at high temperatures the spin modes are described by
independent Gaussian fluctuations. The decay of the correlation
function $G(\mathbf{r},t)$ can then be interpreted as a random walk of the
magnetization through the lattice as in classical diffusion.

It is important to note that this definition of diffusion is not
obviously related to that of diffusive transport given earlier, since
the two definitions refer to two different correlation functions. In
the phenomenological theory, transport is found to be diffusive
because the local magnetization obeys the diffusion equation and the
dc conductivity is therefore finite. However, more generally diffusion
in the autocorrelation function for the density of the conserved
quantity does not exclude the possibility of ballistic transport,
understood as a nonzero  long-time value for the
current-current correlation function.

The applicability of the phenomenological theory of diffusion to spin
chains described by the integrable $XXZ$ model is of course
questionable. Even at high temperatures, the spin dynamics is likely
to be constrained by the nontrivial conservation laws and the
assumption of independent modes is not expected to hold. In the one
case where the long-time behavior of the autocorrelation function can
be calculated exactly, namely the XX model, which is equivalent to
free spinless fermions, diffusion does not occur since $G(x=0,t)\sim
t^{-1}$ for large $t$, as opposed to $t^{-1/2}$ expected for diffusion
in one dimension. For decades, a great deal of effort has been made to
compute the autocorrelation function for the $XXZ$ model with general
values of anisotropy, particularly for the Heisenberg
model.\cite{Carboni,BoehmLeschke,BoehmViswanath,FabriciusLoew,FabriciusMcCoy,StarykhSandvik,
  SirkerDiff} While ED is always limited to small systems and short
times (out to $t\sim 6$ in units of inverse exchange constant for $L=16$
sites\cite{FabriciusMcCoy}), QMC\cite{StarykhSandvik} is plagued by
the analytic continuation and cannot resolve singularities associated
with the long-time behavior. Although the more recently developed
density matrix renormalization group (DMRG) method applied to the
transfer matrix\cite{SirkerKluemperDTMRG} works directly in the
thermodynamic limit, it is also restricted to intermediate times and
has not detected a diffusive contribution at low
temperatures.\cite{SirkerDiff} Nonetheless, these works have concluded
in favor of the existence of diffusion for the Heisenberg model at
high temperatures.

Although diffusion was originally proposed to describe spin dynamics
at high temperatures, the paradigm has been used to interpret nuclear
magnetic resonance (NMR) experiments that measure the spin-lattice
relaxation rate $1/T_1$ of spin chains at low
temperatures.\cite{BoucherBakheit,TakigawaMotoyama96,
  ThurberHunt,KikuchiKurata} Strictly speaking, linear response theory
expresses $1/T_1$ in terms of the Fourier transform of the transverse
spin-spin correlation function $\langle
S^+_j(t)S^-_{j^\prime}(t)\rangle$. The small magnetic field applied in
NMR experiments breaks the rotational spin invariance of Heisenberg
chains from SU(2) down to U(1) and the total $S^\pm$ are not
conserved. However, if the experiments are in the regime of
temperatures small compared to the exchange constant, but large
compared to the nuclear or electronic Larmor frequencies then the
transverse correlation function can be traded for the longitudinal one
calculated for the electronic Larmor frequency. We will discuss this
point in more detail in Section \ref{appD}. Up to $q$-dependent form
factors that stem from the spatial dependence of the hyperfine
couplings, $1/T_1$ is then proportional to the dynamical
autocorrelation $G(\vec{r}=0,\omega)$, with $\omega\propto h$ equal to
the Larmor frequency of an electron in a magnetic field $h$.  If
diffusion is present, with $G(\vec{r}=0,t)\sim t^{-d/2}$ given as in
the phenomenological theory, $1/T_1$ behaves as $1/T_1\propto
G(\vec{r}=0,\omega)\sim \omega^{d/2-1}$ in $d$ dimensions. In
particular, in the one-dimensional case $1/T_1$ diverges at low
frequencies as $ 1/T_1\sim 1/\sqrt{\omega}\sim 1/\sqrt{h}$. This type
of behavior has been observed for gapped $S=1$ spin
chains\cite{TakigawaAsano} and gapless spin chains with large
half-integer $S$.\cite{BoucherBakheit} More surprisingly, spin
diffusion has also been observed in $S=1/2$ chain compounds by
NMR\cite{ThurberHunt,KikuchiKurata} and by muon spin
resonance.\cite{PrattBlundell} It is important to note that in the NMR
experiment of Ref.~[\onlinecite{ThurberHunt}], the $q$-dependence of the
form factor suppresses the contribution from $q\sim \pi$ modes in the
autocorrelation function, so that the $1/T_1$ signal is completely
dominated by $q\sim0$ modes.

The observation of spin diffusion in $S=1/2$ Heisenberg chains is
puzzling from the point of view of the integrability-transport
conjecture. Since the experimental diffusion constant was found to be
fairly large,\cite{ThurberHunt} it becomes important to determine
whether the diffusion constant is mainly determined by
integrability-breaking interactions present in the real system or by
umklapp processes already contained in the integrable Heisenberg
model.

Recently, we have shown using a field theory approach that the
long-time behavior of the autocorrelation function and the transport
properties are directly related and can be obtained from the same
retarded Green's function at low temperatures.\cite{SirkerPereira} The
analytical results were supported by DMRG calculations for the
time-dependent current-current correlation function as well as by a
comparison with the NMR experiment on Sr$_2$CuO$_3$.\cite{ThurberHunt}
We also argued that ballistic transport can be reconciled with
diffusion in the autocorrelation function because ballistic channels
of propagation can coexist with diffusive ones. This can be made
precise with the help of the memory matrix approach,\cite{RoschAndrei}
which allows one to incorporate known conservation laws in the low
energy effective theory.  However, ballistic and diffusive channels
compete for spectral weight of the spin-spin correlation function. Our
field theoretical results for the $XXZ$ chain at $h=0$ -- valid at
finite temperatures small compared to the exchange energy -- are in
very good agreement with the diffusive response measured
experimentally\cite{ThurberHunt} and with time-dependent DMRG results
if we assume that the Drude weight vanishes completely. Although a
small Drude weight at finite temperatures cannot be excluded, a
combination of the numerical data with the memory matrix approach
implies that it has to be smaller than the values obtained in the BA
calculation by Kl\"umper {\it et al.} [\onlinecite{KluemperJPSJ}] and
by QMC calculations\cite{AlvarezGros,AlvarezGros2}. The results in
Ref.~[\onlinecite{SirkerPereira}] were further supported by a recent
QMC study.\cite{GrossjohannBrenig} In the latter work the problems
arising from analytical continuation of numerical data were
circumvented by comparing with the field theory
result\cite{SirkerPereira} transformed to imaginary times.

The purpose of this paper is to provide details of the calculations
for the $XXZ$ chain in Ref.~[\onlinecite{SirkerPereira}]. Furthermore,
we present an extension of these methods to charge transport in the
attractive Hubbard model as well as a discussion of the transport
properties of the Haldane-Shastry chain. Our paper is organized as
follows: In Sec.~\ref{Spin} we study the spin current in the
spin-$1/2$ Heisenberg chain. We discuss the relation between the
current-current and the spin-spin correlation function at low
temperatures, explain in detail how our results relate to previous BA
and QMC calculations, and discuss consequences for electron spin
resonance and the finite-temperature broadening of the dynamic spin
structure factor. In Sec.~\ref{secHS}, we discuss spin transport in
the Haldane-Shastry model and point out that the current operator is a
nonlocal conserved quantity in the thermodynamic limit. In
Sec.~\ref{Charge} we show that many of the results we obtained for the
spin current in the Heisenberg model also directly apply to the charge
current in the attractive Hubbard model.  Finally, we give a summary
and some conclusions in Sec.~\ref{Conclusions}.

\section{The spin current in the spin-$1/2$ $XXZ$ model}
\label{Spin}
The $XXZ$ model (\ref{HXXZspin}) is exactly solvable by Bethe ansatz (BA)
\cite{GiamarchiBook} and for $h=0$ the excitation spectrum is gapless
for $|\Delta|\leq 1$ and gapped for $|\Delta|>1$.

The spin-current density operator is defined from the continuity
equation for the density of the globally conserved spin component \be
\partial_t S^z_l=-i[S^z_l,H]=-(j_{l}-j_{l-1}), \ee which for the $XXZ$
model yields \be 
\label{current_density}
j_l=-\frac{iJ}{2}(S_l^+ S_{l+1}^{-}-S_{l+1}^+
S_{l}^{-}).  \ee For $h\neq 0$, the summed current operator
$\mathcal{J}=\sum_l j_l$ has a finite overlap with the local conserved
quantities of the $XXZ$ model. The simplest nontrivial conserved
quantity is
\begin{widetext}
\be
\mathcal{J}_E=J^2\sum_l\l[S^y_{l-1}S^z_lS^x_{l+1}-S^x_{l-1}S^z_lS^y_{l+1}+\Delta (S^x_{l-1}S^y_lS^z_{l+1}-S^z_{l-1}S^y_lS^x_{l+1})
+\Delta (S^z_{l-1}S^x_lS^y_{l+1}-S^y_{l-1}S^x_lS^z_{l+1})\r].
  \ee 
\end{widetext}
Here $\mathcal{J}_E$ is the energy current operator as obtained from
the continuity equation for the Hamiltonian density for $h=0$
\cite{ZotosPrelovsek,KluemperSakai}. It can be verified that
$\mathcal{J}_E$ is conserved in the strong sense that
$[\mathcal{J}_E,H]=0$. The thermal conductivity therefore only has a
Drude part which can be calculated exactly by BA.\cite{KluemperSakai}
According to Mazur's inequality, Eq.~(\ref{Mazur}), the overlap of
$\mathcal{J}$ with $\mathcal{J}_E$ provides a lower bound for the
Drude weight of the spin conductivity
\begin{equation}
D\geq  D_{\rm Mazur}\equiv\frac{1}{2LT}\frac{\langle
    \mathcal{J}\mathcal{J}_{E}\rangle^{2}}{\langle \mathcal{J}_{E}^{2}\rangle}.
\label{Mazur_exact}
\end{equation}
The advantage of this formula is that -- contrary to Eq.~(\ref{Kubo}) --
it does not require the calculation of dynamical correlation functions
and is thus much more accessible by standard techniques. The
evaluation of (\ref{Mazur_exact}) becomes particularly simple in the
limit $T\to\infty$ leading to\cite{ZotosPrelovsek} 
\be
\label{Mazur_highT}
D_{\rm Mazur}=\frac{J}{T}\frac{4\Delta^{2}m^{2}(1/4-m^{2})}{1+8\Delta^{2}(1/4+m^{2})}\qquad
(T\gg J), \ee where $m=\bra S_l^z\ket$ is the
magnetization. At low temperatures, on the other
hand, standard bosonization techniques can be applied. Furthermore, BA
can be used to evaluate (\ref{Mazur_exact}) for all temperatures as
will be shown in Sec.~\ref{appA}.

All other local conserved quantities can be obtained either
recursively by applying the so-called boost
operator\cite{ZotosPrelovsek} or by taking higher order derivatives of
the quantum transfer matrix of the $XXZ$ Hamiltonian with respect to
the spectral parameter. The local operators obtained this way act  on more and more adjacent sites but they are all even under
particle-hole transformations. As a result, the Mazur bound for the Drude
weight, Eq.~(\ref{Mazur}), vanishes for $h=0$.

\subsection{Low energy effective model and the Mazur bound}
\label{appA}
Bosonization of the $XXZ$ model, Eq.~(\ref{HXXZspin}), in the gapless
regime at zero field leads to the effective Hamiltonian
\cite{GiamarchiBook,EggertAffleck92,Lukyanov} \bea H&=&H_0+H_{\rm
  u}+H_{\rm bc},
\nn\\
\label{Ham_eff}
H_0 &=& \frac{v}{2} \int dx \l[\Pi^2+(\partial_x\phi)^2\r], \nn \\
H_{\rm u}&=&\lambda \int dx
\cos(\sqrt{8\pi K}\phi), \label{HUmklapp}\\
H_{\rm bc}&=&- 2\pi v \lambda_+ \int dx (\partial_x \phi_R)^2(\partial_x \phi_L)^2  \nn \\
&-& 2\pi v \lambda_- \int dx \l[(\partial_x \phi_R)^4 + (\partial_x \phi_L)^4 \nn
\r] \; .
\end{eqnarray}
Here, $H_0$ is the standard Luttinger model and $H_{\rm u}$ and $H_{\rm bc}$
are the leading irrelevant perturbations due to Umklapp scattering and band
curvature, respectively. The bosonic field $\phi=\phi_R + \phi_L$ and its
conjugate momentum $\Pi$ obey the canonical commutation
relation $[\phi(x),\Pi(x')] =\im \delta(x-x')$. The long-wavelength ($q\sim0$)
fluctuation part of the spin density is related to the bosonic field by
$S^z_j\sim \sqrt{K/2\pi}\partial_x\phi$. The spin velocity and the Luttinger
parameter $K$ are known exactly from Bethe ansatz
\begin{equation}
\label{K_and_v}
v=\frac{\pi\sqrt{1-\Delta^2}}{2\arccos\Delta}\quad , \quad
K=\frac{\pi}{\pi-\arccos\Delta} \; .
\end{equation}
In this notation, $K=2$ at the free fermion point ($\Delta=0$) and
$K=1$ at the isotropic point ($\Delta =1$). The amplitudes $\lambda$, $\lambda_+$, and
$\lambda_-$ are also known exactly\cite{Lukyanov}
\begin{eqnarray}
\label{lambdas}
\lambda &=&
\frac{K\Gamma(K)\sin(\pi/K)}{\pi\Gamma(2-K)}\l[\frac{\Gamma\l(1+\frac{1}{2K-2}\r)}{2\sqrt{\pi}\Gamma\l(1+\frac{K}{2K-2}\r)}\r]^{2K-2},
  \nn \\
\lambda_+ &=& \frac{1}{2\pi}\tan\frac{\pi K}{2K-2} ,\\
\lambda_- &=& \frac{1}{12\pi K}
\frac{\Gamma\l(\frac{3K}{2K-2}\r)\Gamma^3\l(\frac{1}{2K-2}\r)}{\Gamma\l(\frac{3}{2K-2}\r)\Gamma^3\l(\frac{K}{2K-2}\r)} .\nn
\end{eqnarray}

In the gapless phase, the Mazur bound can be calculated in the
low-temperature regime using the field theory representations of
$\mathcal{J}$ and $\mathcal{J}_E$. In the continuum limit, the
continuity equation becomes\be
\partial_t S^z(x)+\partial_x j(x)=0.\nonumber \ee Using the bosonized
form for the spin density, we obtain for the effective model
(\ref{Ham_eff})
\bea \mathcal{J}&=& -\sqrt{\frac{K}{2\pi}}\int
dx\,\partial_t \phi \\
&=& -v\sqrt{\frac{K}{2\pi}}\int
dx\l[\Pi-\frac{\pi}{2}(\lambda_++2\lambda_-) \Pi^3\r. \nn \\
&-&\l.\frac{\pi}{2}(-\lambda_++6\lambda_-)\Pi\,(\partial_x\phi)^2\r].\nn
\label{Jwithcurvature}
\eea In the following we neglect the corrections to the current
operator due to band curvature terms and calculate the Mazur bound
$D_{\rm Mazur}$ for the Luttinger model $H_0$ using the current
operator \be \mathcal{J}\approx-v\sqrt{\frac{K}{2\pi}}\int dx\,\Pi \ee
and the energy current operator \be \mathcal{J}_E\approx-v^2\int dx \,
\Pi \,\partial_x\phi.  \ee At zero field, the overlap vanishes because
$\mathcal{J}$ and $\mathcal{J}_E$ have opposite signatures under the
particle-hole transformation $\phi\to-\phi,\Pi\to-\Pi$. A small
magnetic field term in Eq.~(\ref{HXXZspin}), on the other hand, can be
absorbed by shifting the bosonic field (here we set $\mu_B=1$)\be
\phi\to \phi +\frac{h}{v}\sqrt{\frac{K}{2\pi}}x.\label{hshift}\ee In this case, the
conserved quantity becomes \cite{RoschAndrei}\be
\label{conserved_finite_field}
\tilde{\mathcal{J}}_E=-v^2\int dx \, \Pi
\,\partial_x\phi-hv\sqrt{\frac{K}{2\pi}}\int dx \,
\Pi=\mathcal{J}_E+h\mathcal{J}.  \ee We calculate the equal time
correlations within the Luttinger model and find
\begin{equation}
 D_{\rm Mazur}=\frac{1}{2TL}\frac{\langle
    \mathcal{J}\tilde{\mathcal{J}}_{E}\rangle^{2}}{\langle \tilde{\mathcal{J}}_{E}^{2}\rangle}=\frac{vK/4\pi}{1+\frac{2\pi^{2}}{3K}\left(\frac{T}{h}\right)^{2}} \qquad (T,h\ll J).
\label{eq:mazurbound}
\end{equation}
We note that in the limit $T/h\to 0$ the Mazur bound obtained from the
overlap with $\tilde{\mathcal{J}_E}$ saturates the exact zero
temperature Drude weight $D(T=0)=vK/4\pi$ \cite{ShastrySutherland}.

\begin{figure}
\includegraphics*[width=1.0\columnwidth]{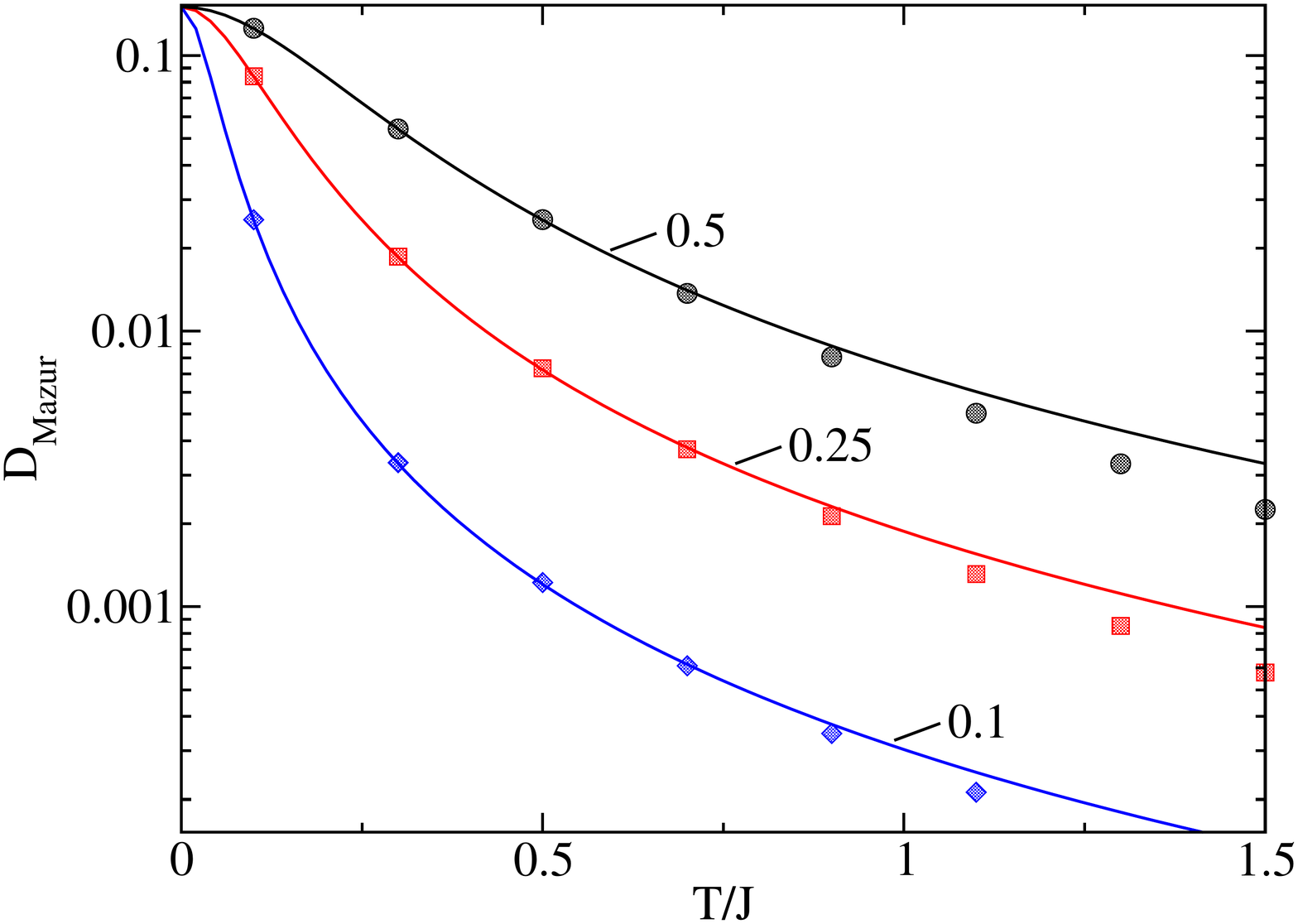}
\caption{Mazur bound $ D_{\rm Mazur}$ as defined in
  (\ref{Mazur_exact}) for $\Delta=\cos(\pi/4)$ and magnetic fields $h$
  as indicated on the plot. The symbols correspond to the exact
  calculation using the Bethe ansatz, the lines represent the field
  theoretical formula (\ref{eq:mazurbound}).}
\label{Fig_Mazur}
\end{figure}

One can also use the Bethe ansatz to calculate the Mazur bound in
Eq.~(\ref{Mazur_exact}) exactly. To do so we computed the equal
time correlations using a numerical solution of the nonlinear integral
equations obtained within the Bethe ansatz formalism of
Refs.~[\onlinecite{SakaiKluemperJPSJ,BortzGoehmann}]. In
Fig.~\ref{Fig_Mazur}, the numerical Bethe ansatz solution for small
values of magnetization $m$ is compared to the field theory formula
(\ref{eq:mazurbound}).
Remarkably, the free boson result in Eq.~(\ref{eq:mazurbound})
fits well the behavior of the exact Mazur bound out to temperatures of
order $J$.

\subsection{Retarded spin-spin correlation function}
\label{appB}
We will now proceed with the field theory calculations in the
following way. We first {\it assume} that the Drude weight is not
affected by unknown nonlocal conserved quantities. In this case we
are left with the following picture based on the Mazur bound and
including all conserved quantities related to integrability: For
finite magnetic field the Drude weight is a continuous function of
temperature. At zero field, however, the Drude weight is only finite
at $T=0$ but drops abruptly to zero for arbitrarily small
temperatures. Within the effective field theory such a possible
broadening of the delta-function peak at finite temperatures has to be
related to inelastic scattering between the bosons which can relax the
momentum. Such a process is described by the Umklapp term in
Eq.~(\ref{Ham_eff}) which we have ignored so far.  We will now include
this term as well as the band curvature terms in a lowest order
perturbative calculation. We want to stress that such an approach does
not know about conservation laws which could protect a part of the
current from decaying. The parameter-free result derived here is
expected to be correct if the Drude weight is indeed zero and allows
to check the validity of the assumption by comparing with experimental
and numerical results.  Importantly, we can also systematically study
how this result is modified if such conservation laws do exist after
all. This case will be considered in section \ref{appF}.

We are interested in the retarded spin-spin correlation function
$\chi_{\rm ret}(q,\omega)$, which can be obtained from the Matsubara
correlation function \be
\chi(q,i\omega_n)=-\frac{1}{N}\sum_{l,l^\prime}^Ne^{-iq(l-l^\prime)}\int_0^{1/T}d\tau\,e^{i\omega_n
  \tau}\bra S_l^z(\tau)S_{l^\prime}^z(0) \ket \ee by the analytic
continuation $i\omega_n\to\omega+i0^+$. We will show that in the
low-temperature limit this correlation function determines {\it both}
the decay of the current-current correlation function as well as the
spin-lattice relaxation rate $1/T_1$. In the low-energy limit, we
follow Ref.~[\onlinecite{OshikawaAffleck}] and relate the
long-wavelength part of the retarded spin-spin correlation function to
the boson propagator
\begin{equation}
\label{SelfE}
\frac{\chi_{\rm ret}(q,\omega)}{Kq^2/2\pi}=\langle\phi\phi\rangle^{\rm ret}(q,\omega)=\frac{v}{\omega^2-v^2q^2-\Pi^{\rm ret}(q,\omega)}.
\end{equation}
We calculate the self-energy $\Pi^{\rm ret}(q,\omega)=\Pi_{\rm u}^{\rm ret}(q,\omega)+\Pi_{\rm bc}^{\rm ret}(q,\omega)$ by perturbation theory to second order in $H_{\rm u}$ and first order in $H_{\rm bc}$. We first focus on the  half-filling case ($h=0$); the 
finite field case is discussed at the end of this sub-section.

The contribution from Umklapp scattering reads
\begin{equation}
\label{SelfE2}
\Pi_{\rm u}^{\rm ret}(q,\omega) = 4\pi K v \lambda^2 \l[F^{\rm ret}(q,\omega) - F^{\rm  ret}(0,0) \r],
\end{equation}
where \cite{Schulz86}
\bea
\label{F_func4}
F^{\rm ret}(q,\omega) &=& -\frac{v}{T^2}\l(\frac{\pi T}{v}\r)^{4K}\sin(2\pi K) \\
&\times& I\left(\frac{\omega+vq}{2T}\right)
I\left(\frac{\omega-vq}{2T}\right),\nn\eea
with \be I(z)=\int_0^\infty \frac{\e^{izu}\, du}{\sinh^{2K}(\pi u)}
=\frac{2^{2K-1}}{\pi }B\l(K-\frac{iz}{2\pi },1-2K\r) ,\label{integral}
\ee where $B(x,y)=\Gamma(x)\Gamma(y)/\Gamma(x+y)$ is the beta
function. For $K>1/2$, we need a cutoff in the integral $I(z)$ in
Eq.~(\ref{integral}). However, the imaginary part of $I(z)$ does
not depend on the cutoff scheme used \cite{Schulz86}. The expansion of
Eq.~(\ref{integral}) for $|\omega\pm vq|\ll T$ yields both a real
and an imaginary part for $\Pi_{\rm u}^{\rm ret}(q,\omega)$.  The
calculation of $\Pi_{\rm bc}^{\rm ret}(q,\omega)$ is also standard. In
contrast to $\Pi_{\rm u}^{\rm ret}(q,\omega)$, the result for
$\Pi_{\rm bc}^{\rm ret}(q,\omega) $ is purely real, as band curvature
terms do not contribute to the decay rate. The end result is\be
\Pi^{\rm ret}(q,\omega) \approx-2i\gamma\omega-b\omega^2+c v^2
q^2.\label{Pi2ndorder} \ee 
In the anisotropic case, $-1<\Delta<1$, the parameters are given by
\begin{eqnarray}
\label{selfE_parameters}
2\gamma &=& Y_1 T^{4K-3} \nn \\
b &=& \underbrace{(Y_2-Y_3) T^{4K-4}}_{b_2} + \underbrace{Y_4T^2}_{b_1} \\
c &=& \underbrace{-(Y_2+Y_3)T^{4K-4}}_{c_2} \underbrace{-Y_4T^2}_{c_1}. \nn 
\end{eqnarray}
Here $b_1$ and $c_1$ ($b_2$ and $c_2$) are the parts stemming from the
band curvature (umklapp) terms, respectively. In
Eq.~(\ref{selfE_parameters}) we have used the following functions
\begin{eqnarray}
\label{selfE_parameters2}
Y_1&=& \Lambda \frac{B(K,1-2K)}{\sqrt{\pi}2^{2K+1}}\cot(\pi K), \nn \\
Y_2 &=& \Lambda \frac{B(K,1-2K)}{\pi^{5/2}2^{2K+4}}(\pi^2-2\Psi'(K)), \nn \\
Y_3 &=& \Lambda \frac{1}{\pi 2^{4K+4}}\cot^2(\pi K) \Gamma(1/2-K)\Gamma(K) ,\\
Y_4 &=& \frac{\pi^2}{6v^2}(\lambda_+ + 6\lambda_-),\nn\\
\Lambda &=& 4\pi K \lambda^2\sin(2\pi
K)\l(\frac{2\pi}{v}\r)^{4K-2}\Gamma(1/2-K)\Gamma(K), \nn
\eea
with $\Psi(x)$ being the Digamma function.

At the isotropic point, $\Delta=1$, Umklapp scattering becomes marginal and can be taken into account by replacing the Luttinger parameter by a running coupling constant, $K\to 1+g(T)/2$. In this case we find
\begin{eqnarray}
\label{selfE_parameters_iso}
2\gamma &=& \pi g^2T ,\nn \\
b &=& \frac{g^2}{4}-\frac{g^3}{32}\l(3-\frac{8\pi^2}{3}\r)+\frac{\sqrt{3}}{\pi}T^2, \\
c &=& \frac{g^2}{4}-\frac{3g^3}{32}-\frac{\sqrt{3}}{\pi}T^2 .\nn \; 
\end{eqnarray}
Following Lukyanov \cite{Lukyanov}, the running coupling constant $g(T)$ is determined by the equation 
\begin{equation}
\label{coup_g}
\frac{1}{g}+\frac{\ln g}{2} = \ln\l[\sqrt{\frac{\pi}{2}}\frac{\e^{1/4+\tilde{\gamma}}}{T}\r],
\end{equation}
where $\tilde{\gamma}$ is the Euler constant.
We remark that a similar calculation was attempted in Ref.~[\onlinecite{FujimotoKawakami}], but there the imaginary part of the self-energy was neglected.


This calculation can be extended to finite magnetic field. Shifting the field $\phi$ as in Eq.~(\ref{hshift}), the Umklapp 
term in Eq.~(\ref{HUmklapp}) becomes $H_{\rm u}=\lambda \int dx \cos[\sqrt{8\pi K} \phi + (2Kh/v)x]$. 
As long as $h \ll T$, it is reasonable to keep this oscillating term in the effective Hamiltonian.  However, in a 
renormalization group treatment, after we lower our momentum cut-off below $h$, it should be dropped. 
Our formula for the self-energy, 
in Eq.~(\ref{SelfE2}) is thus modified to
\bea 
\label{Piret_finite_field}
\Pi_{\rm u}^{\rm ret}(q,\omega) &=& 2\pi K v \lambda^2 \l[F^{\rm ret}(q+2Kh/v,\omega)\r. \nonumber \\
&+& F^{\rm ret}(q-2Kh/v,\omega) \\
&-& \l. F^{\rm  ret}(2Kh/v,0)-F^{\rm  ret}(-2Kh/v,0) \r], \nonumber \eea
where $F^{\rm ret}(q,\omega)$ is given by Eqs.~(\ref{F_func4}) and
(\ref{integral}) as before. As a consequence, the relaxation rate for
$h/T\ll 1$ will now be given by $\gamma \sim
T^{4K-3}[1+\mathcal{O}[(h/T)^2]+\cdots]$ (up to logarithmic corrections
in the isotropic case). In the next section we show that the retarded
current-current correlation function can be obtained at low energies
using the calculated self-energy. We will also discuss what the
shortcomings of the self-energy approach are and show that these
shortcomings can be addressed by taking conservation laws into account
explicitly.

\subsection{Decay of the current-current correlation function}
\label{appF}

The time-dependent current-current correlation function can be written
as\bea
\label{JJfourier}
C(t)&\equiv& \frac{1}{L}\bra\mathcal{J}(t)\mathcal{J}\ket\\
&=&-2\int_{-\infty}^\infty\frac{d\omega}{2\pi}\,\frac{e^{-i\omega
    t}}{1-e^{-\omega/T}} \,{\rm
  Im}\bra\mathcal{J};\mathcal{J}\ket_{\rm ret}(\omega) \nn
, \eea where $\bra\mathcal{J};\mathcal{J}\ket_{\rm
  ret}(\omega)$ is the retarded current-current correlation function.
The latter appears in the Kubo formula for the optical conductivity
(\ref{Kubo}) with $\langle E_{\rm kin}\rangle/L =Kv/2\pi$ and the
current operator given by $\mathcal{J}=-\sqrt{K/2\pi}\partial_t\phi$.
One can easily show that
\begin{equation}
\label{rel1}
\langle\partial_t\phi\partial_t\phi\rangle^{ret}(q,\omega) = -v +\omega^2 \langle\phi\phi\rangle^{ret}(q,\omega)
\end{equation}
leading to 
\begin{equation}
\label{rel2}
\langle \mathcal{J};\mathcal{J}\rangle^{ret}(q,\omega) = -\frac{Kv}{2\pi} +\frac{K}{2\pi}\omega^2
\langle\phi\phi\rangle^{ret}(q,\omega) \; .
\end{equation}
The Kubo formula (\ref{Kubo}) can therefore also be written as
\begin{equation}
\label{Kubo2}
\sigma(q,\omega) = \frac{K}{2\pi}\im\omega \langle
\phi\phi\rangle^{ret}(q,\omega) 
\end{equation}
allowing us to use the results for the boson-boson Green's function
from the previous section.
At zero temperature the irrelevant operators in (\ref{Ham_eff}) can be
ignored and the Drude weight of the $XXZ$ model can be obtained using
the free boson propagator
\begin{equation}
\label{free_prop}
\langle\phi\phi\rangle^{ret}(q,\omega) =\frac{v}{\omega^2-v^2q^2} \; .
\end{equation}  
implying 
\begin{eqnarray}
\label{D0}
 D(T=0)\delta(\omega) &\equiv & \frac{1}{2\pi}\lim_{\omega\to 0}\lim_{q\to 0}
\sigma'(q,\omega) \\
&=& \frac{Kv}{4\pi^2} \textrm{Re}\l[\frac{i}{\omega +\im\epsilon}\r]
=\frac{Kv}{4\pi}\delta(\omega) \nn
\end{eqnarray} 
in agreement with Bethe ansatz.\cite{ShastrySutherland}

If we now turn to finite temperatures we can use the result from the
self-energy approach in Eqs.~(\ref{SelfE}, \ref{Pi2ndorder}) and relation
(\ref{Kubo2}), leading to the optical conductivity \be
\label{opticalcond}
\sigma(q,\omega) = \frac{Kv}{2\pi}\frac{\im\omega}{(1+b)\omega^2-(1+c)v^2q^2+2\im\gamma\omega}
\ee 
with the real part being given by
 \be
\label{fullSigma}
\sigma^\prime(q,\omega)=\frac{Kv\omega}{2\pi}\frac{2\gamma\omega}{[(1+b)\omega^2-(1+c)v^2q^2]^2+(2\gamma\omega)^2}.
\ee 
For $q=0$ we find, in particular, a Lorentzian 
\be
\sigma^\prime(\omega)=\frac{vK}{2\pi}\frac{2\gamma}{[(1+b)\omega]^2+(2\gamma)^2}.
\ee 
As expected, the self-energy approach predicts zero Drude weight
whenever $\gamma(T)$ is nonzero. This result is inconsistent with
Mazur's inequality if there exist conservation laws which protect the
Drude weight. We know that this is the case for any arbitrarily small
magnetic field $h$, while our calculations in sub-section IIB gave a 
non-zero $\gamma$ at non-zero field. Whether or not such a conservation law exists also
for $h=0$ is an open question. We will try to tackle this problem by
first studying how the self-energy result is modified by a
conservation law, followed by a comparison with numerical and
experimental results.

It is possible to accommodate the existence of nontrivial
conservation laws by resorting to the memory matrix formalism of
Ref.~[\onlinecite{RoschAndrei}]. This approach starts from the Kubo
formula for a conductivity matrix $\hat\sigma$ which includes not only
the current operator $\mathcal{J}\equiv \mathcal{J}_1$, but also ``slow
modes'' $\mathcal{J}_n$ ($n\geq2$) which have a finite overlap with
$\mathcal{J}$.
The idea is that if $\bra \mathcal{J}_n(t)\mathcal{J}_n(0)\ket$ is a
slowly decaying function of time, the projection of $\mathcal{J}$ into
$\mathcal{J}_n$ governs the long-time behavior of the current-current
correlation function and consequently dominates the low-frequency
transport. The overlap between $\mathcal{J}$ and the slow modes is
captured by the off-diagonal elements of the conductivity matrix. In
practice, only a small number of conserved quantities is included in
the set of slow modes, but the approach can be systematically improved
since adding more conserved quantities increases the lower bound for
the conductivity \cite{JungRoschlowerbound}.
It is convenient to invert the Kubo formula for the conductivity
matrix using the projection operator method
\cite{JungRoschlowerbound}. We introduce the scalar product between
two operators $A$ and $B$ in the Liouville space\be
(A|B)=\frac{T}{L}\int_0^{1/T}d\tau\,\bra A^\dagger
e^{H\tau}Be^{-H\tau}\ket.  \ee Here, $\mathcal{L}$ is the Liouville
superoperator defined by $\mathcal{L}
\mathcal{J}_n=[H,\mathcal{J}_n]$.  For simplicity, we assumed that all
the slow modes have the same signature under time-reversal symmetry.
The conductivity matrix can be written as \be
\sigma_{nm}(\omega)=i\left(\mathcal{J}_n|(\omega-\mathcal{L})^{-1}|\mathcal{J}_m\right).\label{Kubomtrix}
\ee The conductivity in Eq. (\ref{Kubo}) is the $\sigma_{11}$
component of $\hat\sigma$. The susceptibility matrix $\hat\chi$ is
defined as\be \chi_{nm}=T^{-1}(\mathcal{J}_n|\mathcal{J}_m).  \ee We
denote by \be
\mathcal{P}=1-T^{-1}\sum_{nm}\chi^{-1}_{nm}|\mathcal{J}_n)(\mathcal{J}_m|\ee
the projector \emph{out} of the subspace of slow modes. Using
identities for the projection operator, Eq. (\ref{Kubomtrix}) can be
brought into the form \cite{RoschAndrei}\be
\sigma_{nm}(\omega)=i\{[\omega-\hat{M}\hat\chi^{-1}]^{-1}\hat\chi\}_{nm},\label{sigmanm}
\ee where $\hat{M}$ is the memory matrix given by\be
M_{nm}=T^{-1}(\mathcal{J}_n|\mathcal{L}\mathcal{P}\frac{1}{\omega-\mathcal{P}\mathcal{L}\mathcal{P}}\mathcal{P}\mathcal{L}|\mathcal{J}_m).
\ee It can be shown that if there is an exact conservation law (local
or nonlocal) involving one of the slow modes, the memory matrix has a
vanishing eigenvalue, which then implies a finite Drude weight.

In the following we apply the memory matrix formalism to calculate the
conductivity for the low-energy effective model (\ref{Ham_eff}) at
$h=0$, allowing for the existence of a single conserved quantity $Q$,
$[Q,H]=0$. 
The conductivity matrix is then two-dimensional. We choose
$\mathcal{J}_1=\mathcal{J}$ and $\mathcal{J}_2=Q_\perp\equiv
Q-\mathcal{J}(\mathcal{J}|Q)(\mathcal{J}|\mathcal{J})^{-1}$ so that
$\hat\chi$ is diagonal.  At low temperatures $T\ll J$, we can use the
current operator in Eq.~(\ref{Jwithcurvature}); to first order in
$\lambda_\pm$, we obtain \be \chi_{11}\approx
\frac{\bra\mathcal{J}^2\ket}{LT}\approx \frac{vK}{2\pi}(1-b_1), \ee
where $b_1$ is defined in Eq.~(\ref{selfE_parameters}) and we have
used the fact that due to the vanishing of the superfluid density we
have $\langle E_{\rm kin}\rangle/L\approx\langle
\mathcal{J}^2\rangle/LT$. Likewise, $\chi_{22}=(\bra Q^2\ket-\bra
QJ\ket^2/\bra J^2\ket)/(LT)$ can be calculated within the low energy
effective model once a conserved quantity $Q$ has been identified. The
remaining approximation is in the calculation of the memory matrix to
second order in Umklapp. This is analogous to the calculation of the
self-energy $\Pi^{\rm ret}_{\rm u}$ in Eq.~(\ref{SelfE2}). Using the
conservation law, we can write\be \hat{M}\approx M_{11}(\omega)\left(
  \begin{array}{cc}
    1 & -r  \\
    -r & r^2 \end{array} \right), \ee where $r=\bra
Q\mathcal{J}\ket/\bra\mathcal{J}^2\ket$ and \be M_{11}(\omega)\approx
\frac{vK}{2\pi}\frac{\Pi_{\rm u}(\omega)}{\omega}\approx
\frac{vK}{2\pi}(-b_2\omega -2i\gamma),\label{sigmammat} \ee with $b_2$
as given in Eq.~(\ref{selfE_parameters}). Thus, from equation
(\ref{sigmanm}), we find\be \sigma(\omega)=\chi_{11}\l[ \frac{y}{1+y}
\frac{i}{\omega}+\frac{1}{1+y}\frac{i}{\omega-(1+y)\chi_{11}^{-1}M_{11}(\omega)}
\r],
\label{sigmammat2}
\ee where\be y=\frac{r^2\chi_{11}}{\chi_{22}}=\frac{\langle
  \mathcal{J} Q\rangle^2}{\langle \mathcal{J}^2\rangle \langle
  Q^2\rangle -\langle \mathcal{J} Q\rangle^2 }.  \ee For finite
magnetic field, a conserved quantity is given by
$Q=\mathcal{J}_E+h\mathcal{J}$ (see
Eq.~(\ref{conserved_finite_field})) and thus $y\sim (h/T)^2$ in this
case. Equating Eq.~(\ref{opticalcond}) for $q=0$ and
(\ref{sigmammat2}), we find that the memory matrix approach is
equivalent to adopting the self-energy \be \Pi(\omega)\approx
\frac{\omega
  \chi_{11}^{-1}M_{11}(\omega)-b_1\omega^2}{1-y\chi_{11}^{-1}M_{11}(\omega)/\omega}=\frac{-b\omega^2-2i\gamma\omega}{1-y\chi_{11}^{-1}M_{11}(\omega)/\omega}.\label{selfememoryfunction}
\ee As expected, the memory matrix result reduces to the self-energy
result for $y\to 0$. The difference between the self-energy and the
memory matrix result is of higher order in the Umklapp interaction.
Therefore, the conservation law is not manifested in the lowest-order
calculation of the self-energy. Although
Eq.~(\ref{selfememoryfunction}) is not correct beyond
$O(\lambda^2,\lambda_\pm)$, it suggests that the memory matrix
approach corresponds to a partial resummation of an infinite family of
Feynman diagrams which changes the behavior of $\Pi(\omega)$ in the
limit $\omega\to 0$ from $\Pi\to -2i\gamma\omega$ to $\Pi\to
-y^{-1}\omega^2$.

It follows from Eq.~(\ref{Kubo}) and Eq.~(\ref{sigmammat2}) that \bea
\label{ImJJret}
{\rm Im}\bra\mathcal{J};\mathcal{J}\ket_{\rm ret}(\omega) &=&
-\omega{\rm Re
}\,\sigma(\omega)\nonumber \\
&=& -\omega\frac{vK}{2\pi}\l[\frac{\pi
  y(1-b_1)}{1+y}\,\delta(\omega)\r. \\
&+&\l.
\frac{2\gamma}{(1+b_1+b_2^\prime)^2\omega^2+(2\gamma^\prime)^2}\r],
\nonumber \nn \eea where $b_2^\prime=(1+y)b_2$ and
$\gamma^\prime=(1+y)\gamma$. The first term on the right hand side of
Eq.~(\ref{ImJJret}) can be associated with the ballistic channel and
the second one with the diffusive channel. The calculation is valid
also in the finite field case if $h/T\ll 1$ with $y\sim(h/T)^2$ as
already discussed below Eq.~(\ref{sigmammat2}). This means that we
obtain a correction of the relaxation rate in the memory matrix
formalism which is exactly of the same order as obtained previously
from the self-energy approach in this limit (see
Eq.~(\ref{Piret_finite_field})). However, now we see that the weight
is transferred accordingly from the diffusive into a ballistic
channel, a fact which is missed in the self-energy approach.

Substituting Eq.~(\ref{ImJJret}) into Eq.~(\ref{JJfourier}) we can now
evaluate the integral. From the second term in (\ref{ImJJret}) we
obtain an integrand which has poles at frequencies $\omega=\pm 2\im
\gamma^\prime/(1+b_1+b_2^\prime)$ and $\omega=2\pi\im Tn$ with
$n\in\mathbb{Z}$. Since $2\gamma^\prime\ll 2\pi T$ the contributions
of the latter poles can be ignored at times $t\gg (2\pi T)^{-1}$
leading to \be
\label{Ct_full}
C(t)=\frac{Kv}{2\pi(1+y)}\l[yT(1-b_1)-\frac{2\im\gamma''}{1+b_1+b_2^\prime}\frac{\e^{-2\gamma''
    t}}{1-\e^{2\im\gamma''/T}}\r] \ee with $\gamma''
=\gamma'/(1+b_1+b_2')$. We can further expand the denominator in
powers of $\gamma''/T\ll 1$. The first order contribution is real and
given by \be 
\label{Ct_real}
C(t)\sim \frac{vKT}{2\pi(1+y)}\l[y(1-b_1)+\frac{e^{-2\gamma'
    t}}{1+b_1+b_2^\prime}\r].  \ee 
The imaginary part of the correlation function is obtained in second
order in the expansion and is thus suppressed by an additional power
of $\gamma'/T$. From Eq.~(\ref{Ct_real}) we see that in the limit
$t\to\infty$ the current-current correlation function approaches the
value\be \lim_{t\to\infty}C(t)=\frac{vKTy(1-b_1)}{2\pi(1+y)}\approx
\frac{\bra\mathcal{J}^2\ket
  y}{L(1+y)}=\frac{\bra\mathcal{J}Q\ket^2}{L\bra Q^2\ket}, \ee
consistent with the Mazur bound for the Drude weight. For intermediate
times $(2\pi T)^{-1}\ll t\ll 1/\gamma^{\prime}$, we obtain the linear
decay \be
\label{lin_decay}
C(t)\approx \frac{vKT}{2\pi(1+b)}(1-2\gamma t),
\ee independent of $y$ if $b_1,b_2^\prime\ll 1$. Therefore a small
Drude weight cannot be detected in this intermediate time range.

\subsection{Comparison with Bethe ansatz and numerical data for the current-current correlation function}
In the previous section we have derived a result for the optical
conductivity, Eq.~(\ref{ImJJret}), and for the real part of the
current-current correlation function, Eq.~(\ref{Ct_real}), at low
temperatures by a memory-matrix formalism. In this approach we have
explicitly taken into
account the possibility of a nonlocal conservation law -
leading to a nonzero Drude weight at finite temperatures. By comparing our results with the Bethe ansatz
calculations\cite{Zotos,KluemperJPSJ} and numerical calculations of
$C(t)$ we will show that a diffusive channel for transport does indeed
exist. Furthermore, we will try to obtain a rough bound for how large $y$
and therefore the Drude weight $D(T)$ can possibly be.

A finite Drude weight at finite temperatures has been obtained in two
independent Bethe ansatz calculations\cite{Zotos,KluemperJPSJ},
however, the obtained temperature dependence is rather different. In
both works the finite temperature Kohn formula\cite{CastellaZotos},
which relates the Drude weight and the curvature of energy levels with
respect to a twist in the boundary conditions, is used. While Zotos
[\onlinecite{Zotos}] uses a TBA approach based on magnons and their
bound states, Kl\"umper {\it et al.} [\onlinecite{KluemperJPSJ}] use
an approach based on a spinon and anti-spinon particle basis. Both
approaches have been shown to violate exact relations at high
temperatures and are therefore not exact solutions of the problem. The
reason is that within the Bethe ansatz both approaches use assumptions
which have been shown to work in the thermodynamic limit for the
partition function. This, however, does not seem to be the case
for the curvature of energy levels relevant for the Drude weight.
Nevertheless, this does not exclude that these results become
asymptotically exact at low temperatures and arguments for such a
scenario have been given in Ref.~[\onlinecite{KluemperJPSJ}].

To investigate this possibility we start by comparing in
Fig.~\ref{Comp_BA} the Bethe ansatz results from
Ref.~[\onlinecite{KluemperJPSJ}] at low temperatures with the Drude
weight \be
\label{BA_Drude}
D(T) = \frac{Kv}{4\pi(1+b)} \ee obtained by setting $\gamma=0$ in
Eq.~(\ref{opticalcond}).  We see from Eq.~(\ref{Ct_real}) that this
corresponds to the case $y\to\infty$, i.e., in this case there is only
a ballistic channel.
\begin{figure}
\includegraphics*[width=1.0\columnwidth]{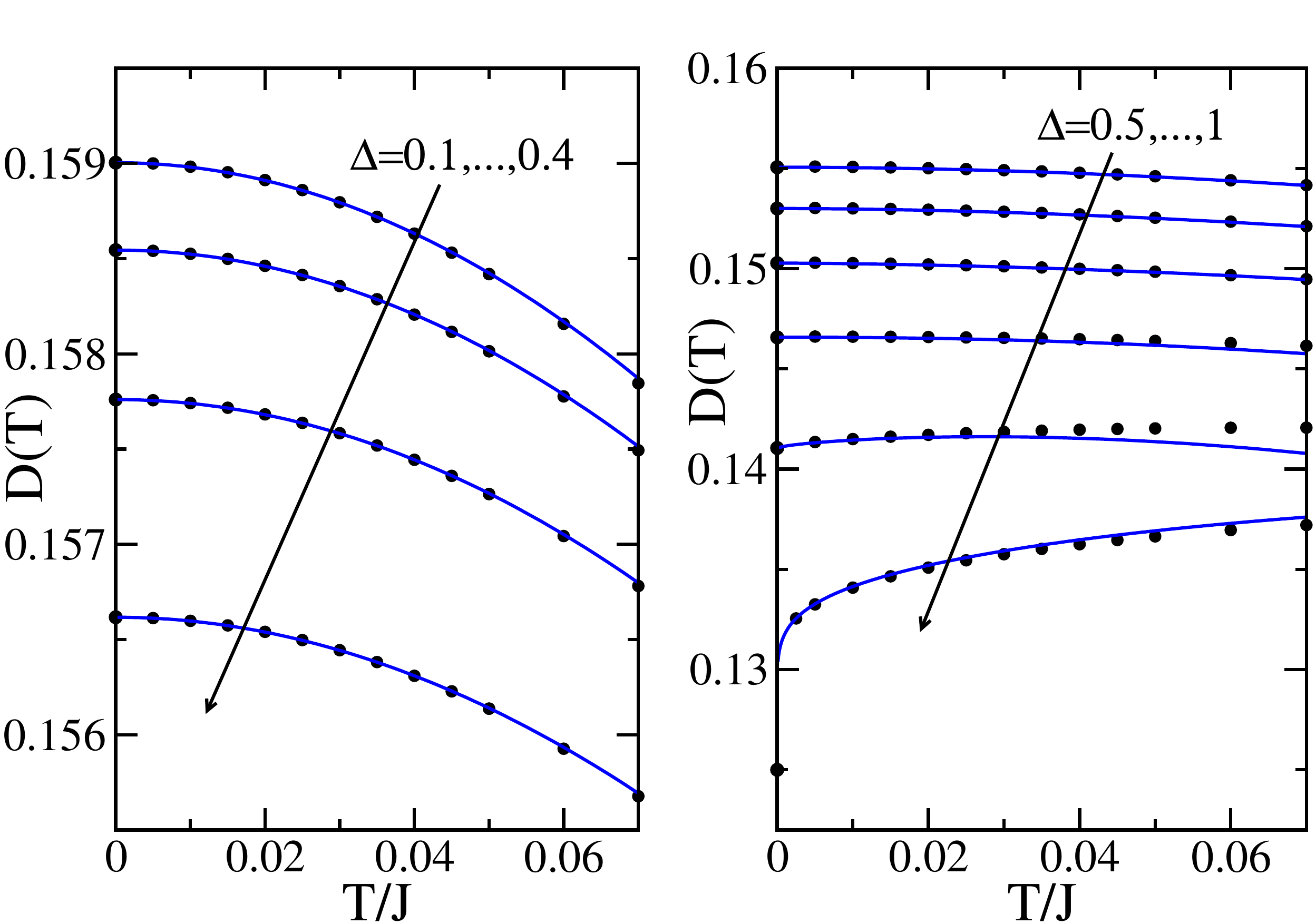}
\caption{The Bethe ansatz results from
  Ref.~[\onlinecite{KluemperJPSJ}] (dots) compared to the field theory
  formula (\ref{BA_Drude}) (lines) obtained by setting $\gamma=0$ by
  hand in Eq.~(\ref{opticalcond}). The $\Delta$ values are indicated
  on the plot.}
\label{Comp_BA}
\end{figure}
Doing so we obtain excellent agreement. From this we draw two
conclusions: First, this BA calculation predicts that even at finite
temperatures the transport is purely ballistic. Second, the
temperature dependent parameter $b$ which we obtained from field
theory in first order in band curvature and second order in Umklapp
scattering is consistent with this BA approach. We note that the case
$\Delta=1/2$ ($K=3/2$) is particularly interesting because here the
contributions $b_1$ from band curvature and $b_2$ from Umklapp
scattering in Eq.~(\ref{selfE_parameters}) both yield a $T^2$
contribution with diverging prefactors. These divergencies cancel
leading to a Drude weight
\be
\label{Drude3.2}
D(T)=\frac{9\sqrt{3}}{32\pi}\frac{1}{1+\tilde b T^2} \ee with $\tilde
b = [142+24\tilde\gamma+48\ln 2+ 60\ln 3 - 24 \ln T -21\zeta(3)]/243$
where $\tilde\gamma$ is the Euler constant and $\zeta$ the Riemann
zeta function.  As is also the case for other thermodynamic
quantities\cite{SirkerBortzJSTAT} we see that the coinciding scaling
dimensions of the terms stemming from Umklapp scattering and band
curvature lead to a term $\sim T^2\ln T$ at this special point. For
$\Delta<1/2$ ($K>3/2$) band curvature gives the dominant temperature
dependence while Umklapp scattering dominates for $\Delta>1/2$
($K<3/2$). For $K>5/4$ ($\Delta\gtrsim 0.81$) a $T^{8K-8}$ term, which
arises in 4th order perturbation theory in Umklapp scattering and
which is not included in our calculations, becomes more important than
the $T^2$ term from band curvature. Therefore the agreement for
$\Delta=0.9$ in Fig.~\ref{Comp_BA} is not quite as good as for the
other values. For $\Delta=1$, Umklapp scattering becomes marginal
leading to a logarithmic temperature dependence of the Drude weight
(\ref{BA_Drude}) with $b$ as given in (\ref{selfE_parameters_iso}).

To see whether or not such a large Drude weight as predicted by
Kl\"umper {\it et al.} is possible and to discuss the second BA
approach by Zotos we now turn to a numerical calculation of $C(t)$.
A dynamical correlation function at finite temperatures can be
obtained by using a density matrix renormalization group algorithm
applied to transfer matrices
(TMRG).\cite{SirkerKluemperDTMRG,SirkerDiff,SirkerPereira} This
algorithm uses a Trotter-Suzuki decomposition to map the 1D quantum
model onto a 2D classical model. For the classical model a so-called
quantum transfer matrix can be defined which evolves along the spatial
direction and allows one to perform the thermodynamic limit exactly.
In order to calculate dynamical quantities, a complex quantum transfer
matrix is considered with one part representing the thermal density
matrix and the other part the unitary time evolution operator. By
extending the transfer matrix one can either lower the temperature
(imaginary time) or increase the real time interval for the
correlation function. The calculation of the current-current
correlation function is particularly complicated because it involves
the summation of all the local time-dependent correlations \be
\label{TMRG_current}
\frac{1}{L}\langle\mathcal{J}(t)\mathcal{J}(0)\rangle = \sum_l \langle
j_l(t) j_0(0)\rangle \ee with the current density $j_l$ as given in
Eq.~(\ref{current_density}). In Fig.~\ref{Fig_TMRG1} the local
correlations $\langle j_l(t)j_0(0)\rangle$ are exemplarily shown for
the case $\Delta=0.6$ and $T/J=0.2$.
\begin{figure}[ht]
\includegraphics*[width=1.0\columnwidth]{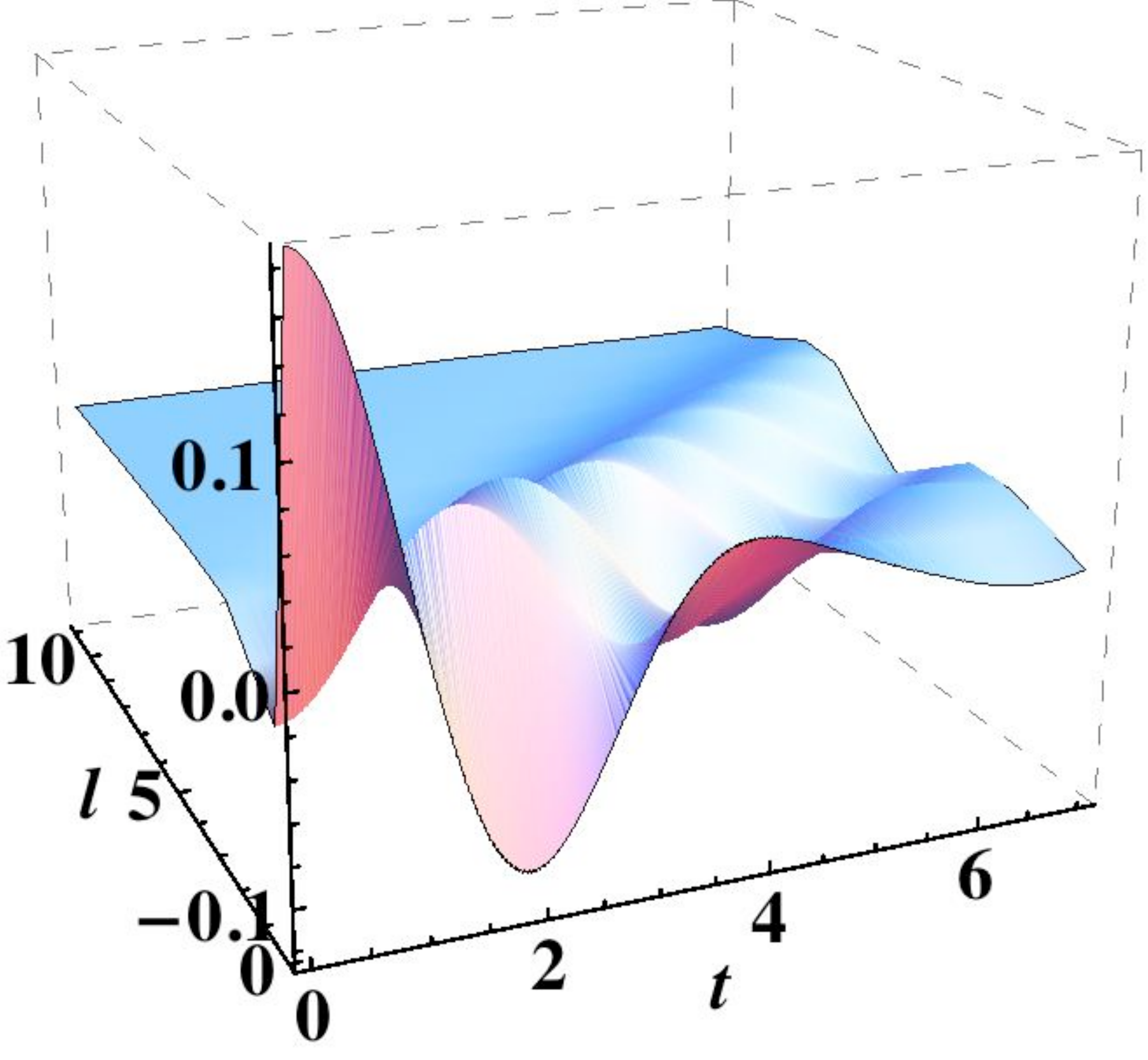}
\caption{$\langle j_l(t)j_0(0)\rangle$ for $\Delta=0.6$ at $T/J=0.2$.}
\label{Fig_TMRG1}
\end{figure} 
In order to obtain converged results, the two-point correlations for
distances up to $l\sim 30$ have to be summed up. This is not possible
using exact diagonalization which is restricted to considerably
smaller system sizes. A good check is obtained by considering the free
fermion case $\Delta=0$. Here $\langle j_l(t)j_0(0)\rangle$ can be
calculated exactly and is non-trivial, however, $\mathcal{J}=\sum_l j_l$
commutes with the Hamiltonian leading to (\ref{TMRG_current}) being a
constant.

We now discuss the same parameter set $\Delta=0.6$ and $T/J=0.2$ as
used above in more detail. In Fig.~\ref{Fig_TMRG2} the real and
imaginary parts of $C(T)/2JT$, obtained in a TMRG calculation with
a $1000$ states per block kept, are shown. Because the imaginary part
is very small, an extrapolation in the Trotter parameter was necessary
restricting the calculations to smaller times than for the real part.
\begin{figure}[ht]
\includegraphics*[width=1.0\columnwidth]{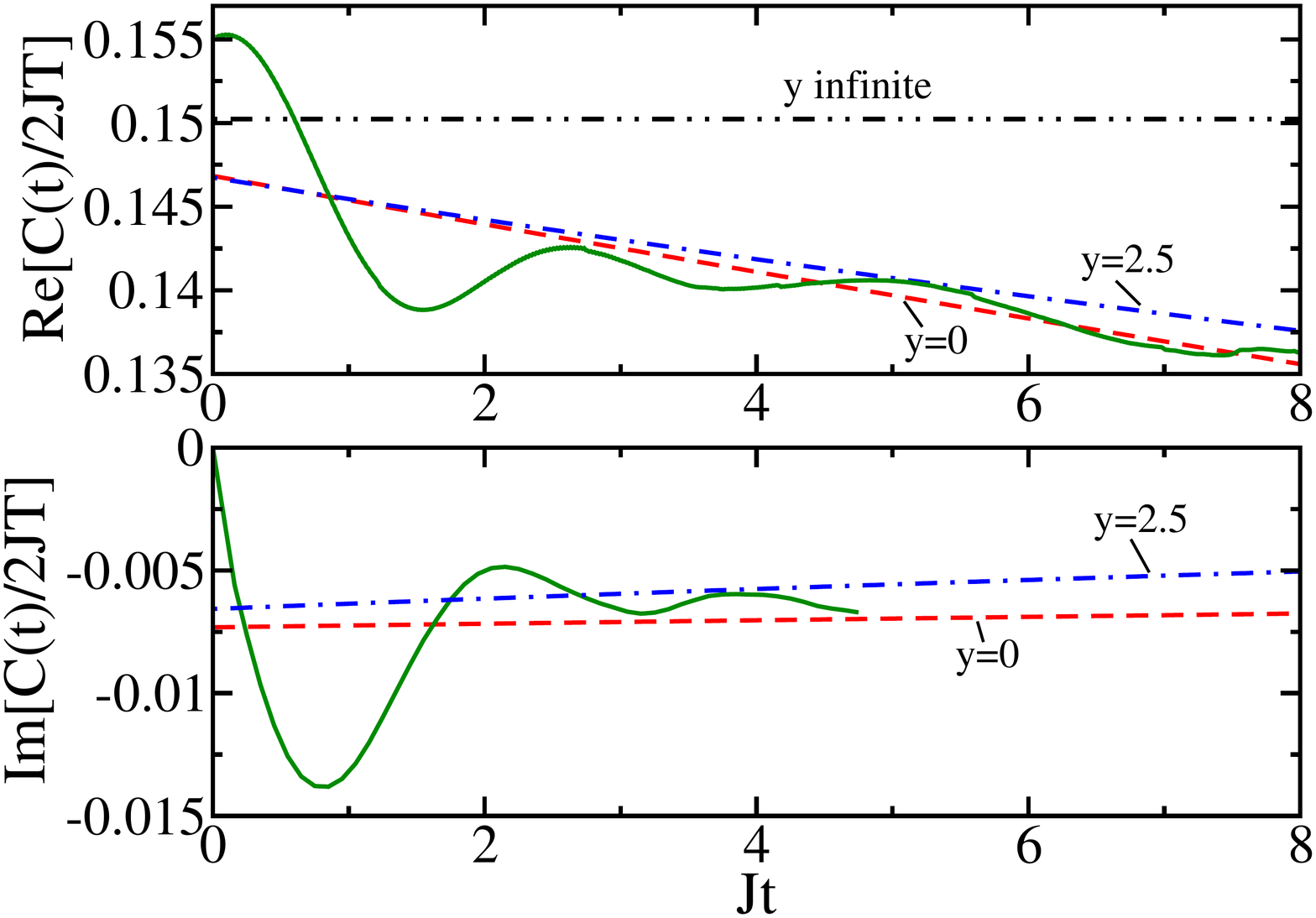}
\caption{Real and imaginary parts of the current correlation function
  for $\Delta=0.6$ and $T/J=0.2$. The dashed, dot-dashed and dotted
  lines correspond to formula (\ref{Ct_full}) with $y$ values as
  indicated on the plot.}
\label{Fig_TMRG2}
\end{figure} 
In the limit $t\to\infty$, the real part would directly yield the
Drude weight (possibly zero). As already discussed, the BA result by
Kl\"umper {\it et al.}  corresponds to purely ballistic transport,
$y\to\infty$. From Fig.~\ref{Fig_TMRG2} we see that this is not
consistent with the numerical data. The BA calculation by Zotos, on
the other hand, predicts a Drude weight $\sim 0.105$ which requires
$y\sim 2.5$. While formula (\ref{Ct_full}) with $y=0$ seems to fit the
numerical results best, we would need to be able to simulate slightly
longer times (a factor of $1.5-2$ should be sufficient) to clearly
distinguish between $y=0$ and $y=2.5$. Most importantly, however, the
numerical data clearly demonstrate that the decay rate $\gamma$ is
nonzero.

In addition to the zero field case, we have also calculated $C(t)$ at
relatively large magnetic fields and various temperatures. As shown in
Fig.~\ref{Fig_TMRG_h1} we find that in such cases $C(t)$ appears to
converge to a finite value within fairly short times.
\begin{figure}
\includegraphics*[width=1.0\columnwidth]{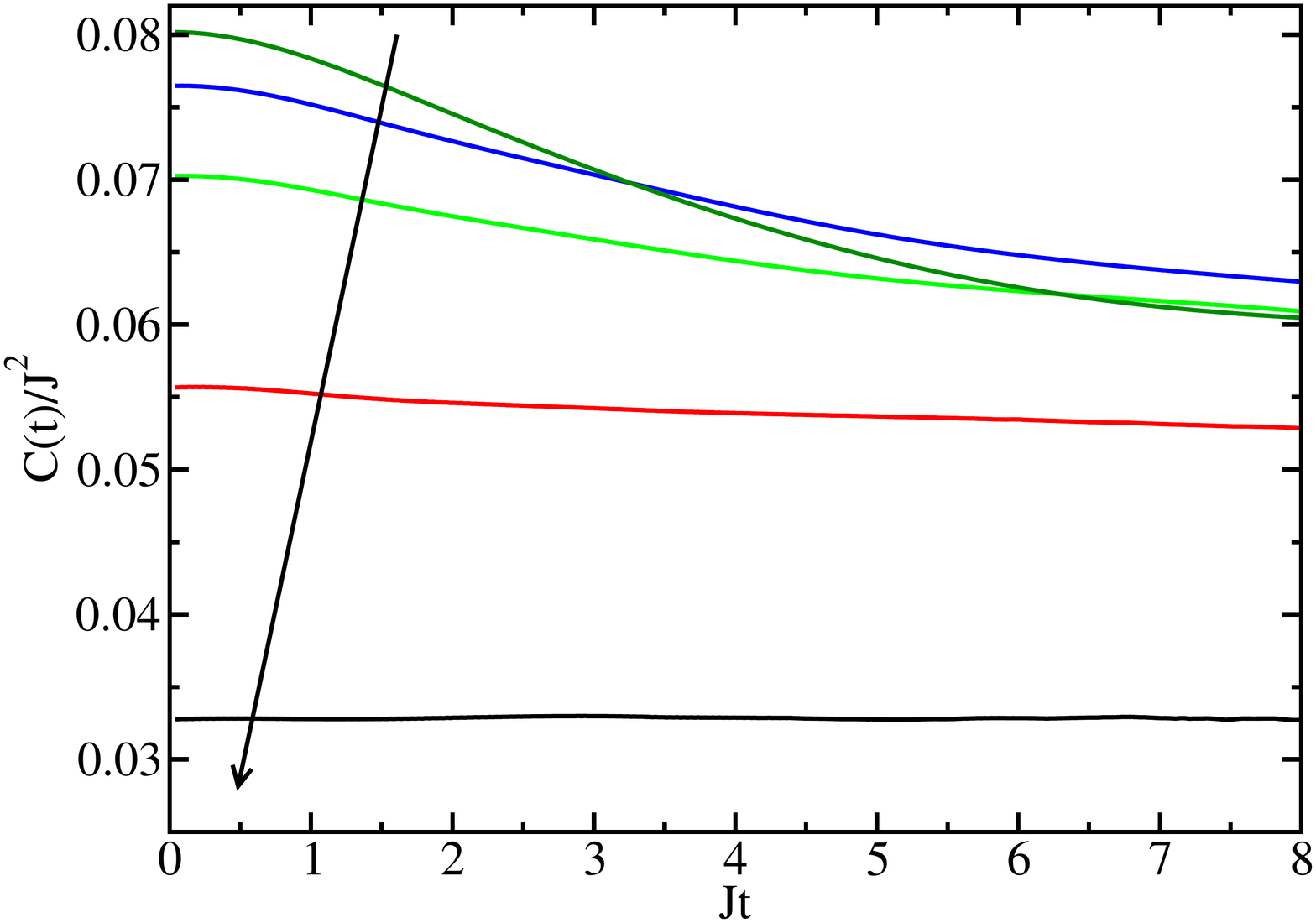}
\caption{$C(t)$ for $\Delta=0.4$ and a magnetization $\langle m\rangle
  =0.3$ at temperatures $T/J=20,1,0.\bar{6},0.4,0.2$ (in arrow
  direction). At low temperatures the asymptotic value seems to be
  reached almost instantaneously. At high temperatures the data seem
  to be consistent with a simple exponential decay to a finite value
  without oscillations. Using an exponential fit and extrapolating to
  infinite temperature we find $C(T\to\infty,t\to\infty) \approx
  0.058$. On the other hand, the Mazur bound, Eq.~(\ref{Mazur_highT}),
  yields $2TD_{\mbox{Mazur}}=0.013$.}
\label{Fig_TMRG_h1}
\end{figure}
Furthermore, we find (see Fig.~\ref{Fig_TMRG_h2}) that at large $\Delta$ and
large magnetic fields the Mazur bound (\ref{Mazur_highT}) almost completely
exhausts the Drude weight at high temperatures. This is consistent with the
findings in Ref.~[\onlinecite{ZotosPrelovsek}] which were based on exact
diagonalization.
\begin{figure}
  \includegraphics*[width=1.0\columnwidth]{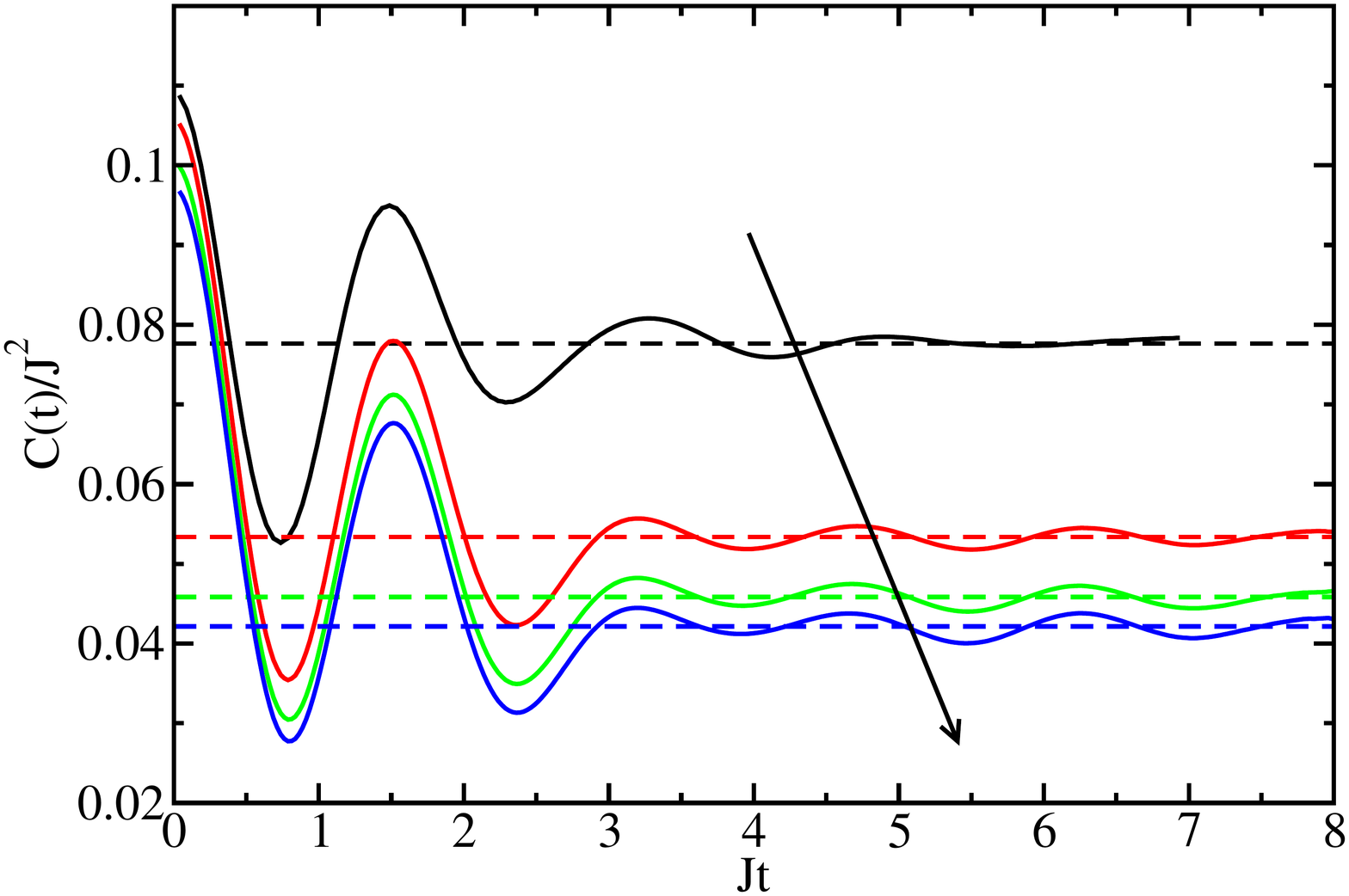}
  \caption{$C(t)$ for $\Delta=4.0$ and a magnetization $\langle
    m\rangle =0.25$ with $T/J=1,5,10,20$ (in arrow direction). In
    addition, linear fits of the data in the regime $Jt\geq 4$ are
    shown (dashed lines). Extrapolating these values we find
    $C(T\to\infty,t\to\infty) \approx 0.0385$. In this case the Mazur
    bound, Eq.~(\ref{Mazur_highT}), yields $2TD_{\mbox{Mazur}}=0.0366$
    for $T\to\infty$ and exhausts 95\% of the Drude weight.}
\label{Fig_TMRG_h2}
\end{figure}

\subsection{Comparison with Quantum Monte Carlo}
\label{appG}
Quantum Monte Carlo (QMC) calculations are performed in an imaginary
time framework. In Refs.~[\onlinecite{AlvarezGros,AlvarezGros2}] the
optical conductivity $\sigma(q,\im \omega_n)$ at Matsubara frequencies
$\omega_n=2\pi T n$, $n\in\mathbb{N}$ has been determined. In order to
answer the question whether or not a finite Drude weight exists at
finite temperatures, one has to perform first the limit $q\to 0$ and
then try to extrapolate in the discrete Matsubara frequencies
``$\omega_n\to 0$''. Doing so the results obtained in
Refs.~[\onlinecite{AlvarezGros,AlvarezGros2}] have been interpreted as
being consistent with the Drude weight found in the BA calculations by
Kl\"umper {\it et al.}~[\onlinecite{KluemperJPSJ}].

By replacing $\omega\to \im\omega_n$, the field theory formula
(\ref{opticalcond}) yields a prediction for $\sigma(q,\im \omega_n)$
which can be directly compared with the QMC results (see Fig.~\ref{Fig_QMC}). 
\begin{figure}
\includegraphics*[width=1.0\columnwidth]{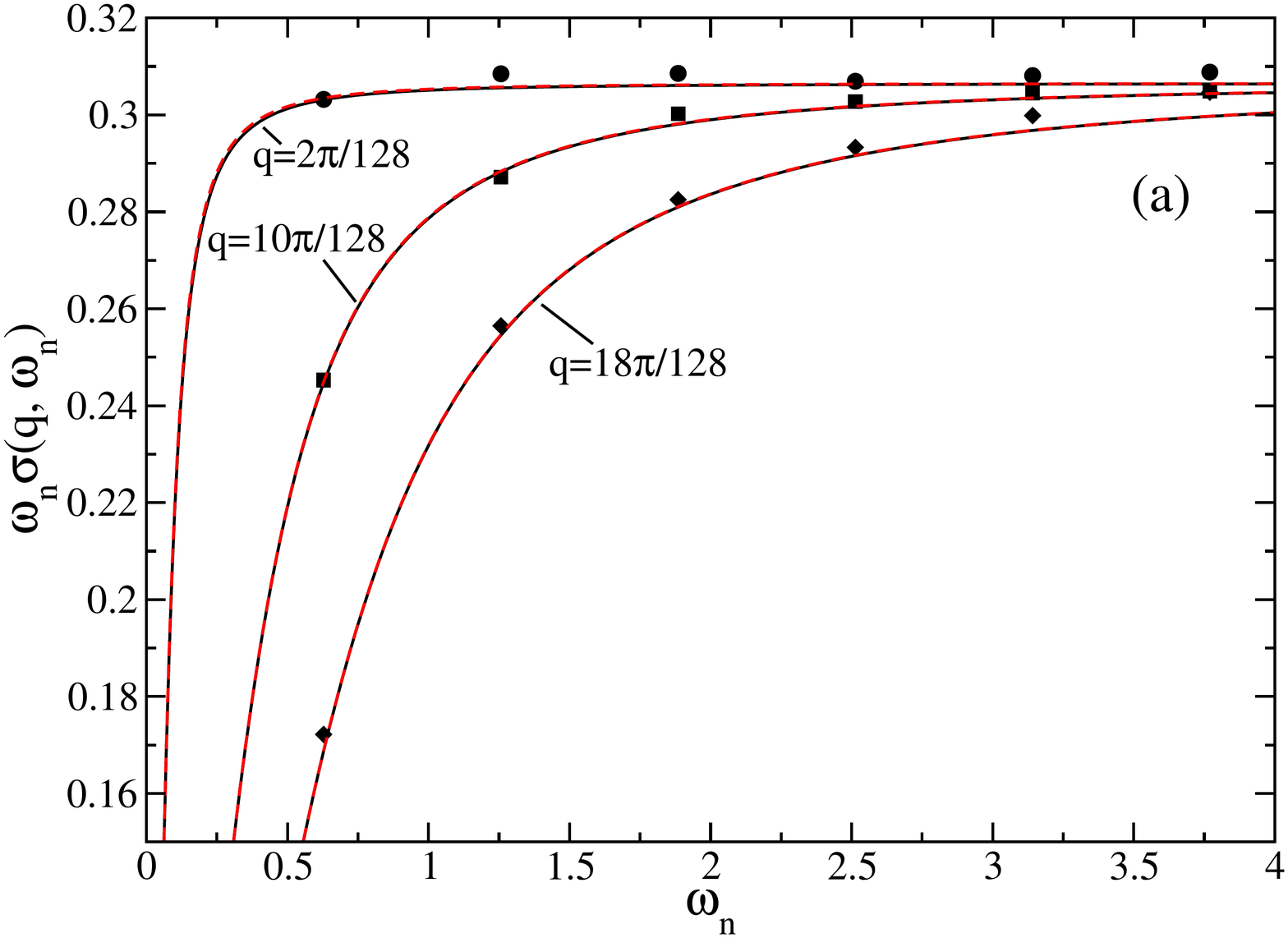}
\includegraphics*[width=1.0\columnwidth]{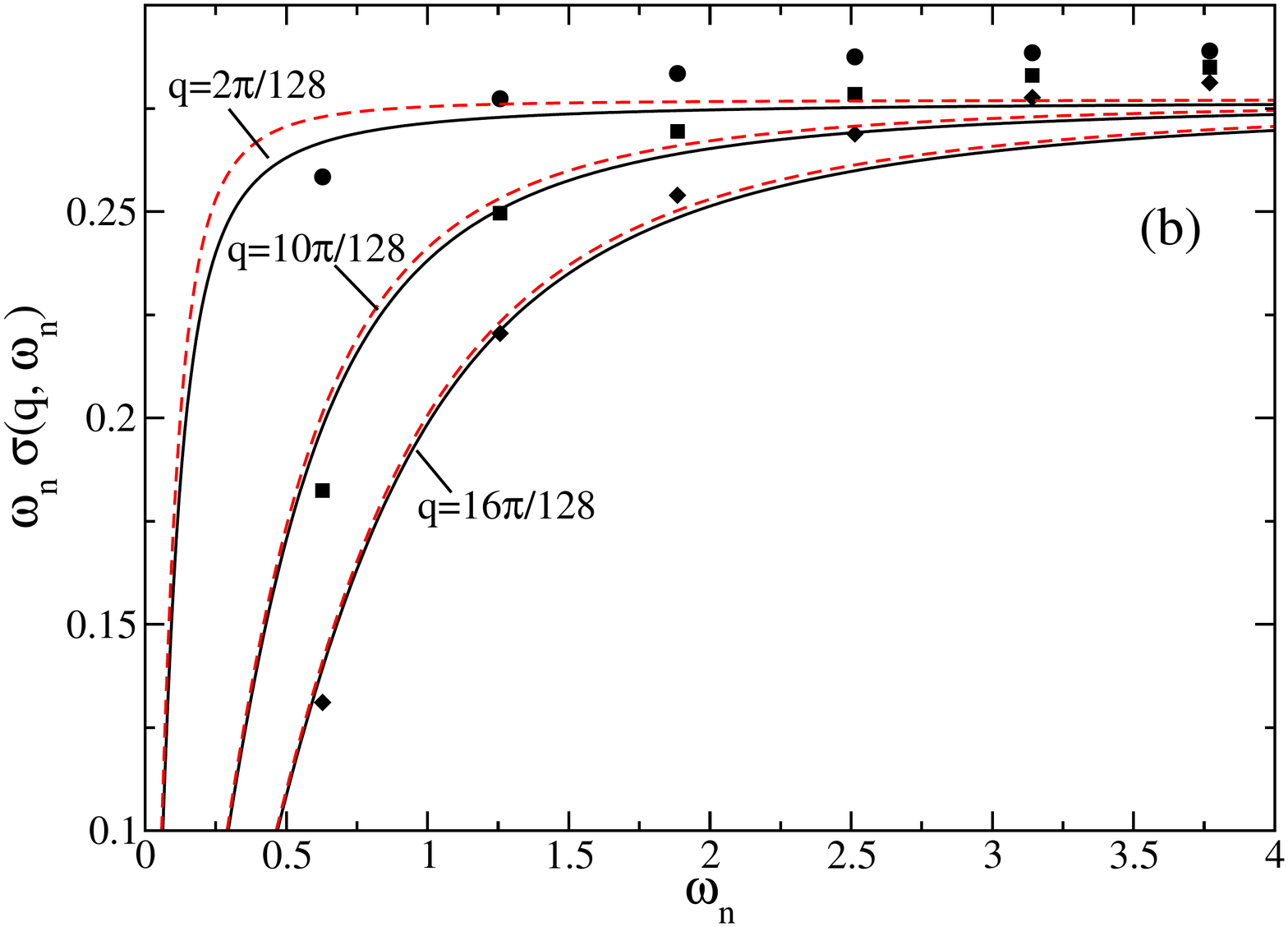}
\caption{QMC data (symbols) for $T/J=0.1$ and a system size $L=128$
  for (a) $\Delta=1/2$, and (b) $\Delta=1$.\cite{AlvarezGros_private}
  In comparison, the field theory result in Eq.~(\ref{opticalcond})
  (solid lines) and the result obtained by setting $\gamma=0$ (dashed
  lines) is shown. For $\Delta=1/2$ both cases are almost
  indistinguishable. For $\Delta=1$ and small $q$ the effect of
  $\gamma$ is largest, however, in this case also the finite size
  corrections are large so that a detailed analysis seems very
  difficult.}
\label{Fig_QMC}
\end{figure}
In addition we also show the result that is obtained by setting
$\gamma=0$ in Eq.~(\ref{opticalcond}), corresponding to the BA
solution in Ref.~[\onlinecite{KluemperJPSJ}].  Since the agreement is
good for both $\gamma=0$ and $\gamma\neq0$ as given by
Eq.~(\ref{selfE_parameters}), we conclude that these QMC calculations
are not of sufficient accuracy to decide whether or not the relaxation
rate vanishes in the integrable model. A general problem in QMC
calculations is that a relaxation rate much smaller than the
separation between Matsubara frequencies, $\gamma\ll
\omega_{n+1}-\omega_n=2\pi T$, cannot be resolved.

Promising seems to be a study of the Heisenberg point, where
$\gamma(T)\sim T/\ln^2(J/T)$ is largest. Indeed, evidence for a
nonzero relaxation rate for this case has been found in a very recent
QMC study by directly comparing with our result (\ref{opticalcond})
transformed to Matsubara frequencies.\cite{GrossjohannBrenig} The
magnitude of $\gamma$ seemed to be roughly consistent with the value
in Eq.~(\ref{selfE_parameters_iso}). However, as has been already
shown in Ref.~[\onlinecite{SirkerPereira}] by comparing with numerical
results for the correlation function $C(t)$, logarithmic correction at
the isotropic point limit the temperature range where the field theory
results are applicable. In QMC calculations a further problem at
$\Delta=1$ is the slow logarithmic decay of finite size corrections.
It would therefore be desirable to perform a similar study for
$\Delta\sim 0.6-0.8$ where $\gamma$ is still fairly large and the
field theory seems to work well up to temperatures of the order
$T/J\sim 0.2$ as we demonstrated in the previous section and in
Ref.~[\onlinecite{SirkerPereira}].

\subsection{Spin diffusion}
\label{appC}
So far we have used our result for the self-energy of the bosonic
propagator to discuss the transport properties of the spin chain. In
this and the following sections we will discuss diffusive properties
characterized by the long-time behavior of the spin-spin correlation
function $\bra S_{l+x}^z(t)S^z_l(0)\ket$. We concentrate again on the
case of zero field.  For $T=0$, it is known\cite{PereiraWhite} that
the slowest decaying term in the autocorrelation function for
$0<\Delta<1$ is of the form \be \bra S_{l}^z(t)S^z_l(0)\ket \sim
\frac{e^{-iWt}}{t^\eta},\quad (T=0) \ee with $W=v$ and $\eta=(K+1)/2$.
This oscillating term is attributed to $q=\pi/2$ high-energy
particle-hole excitations with a hole near the bottom of the band and
a particle at the Fermi surface, or a particle at the top of the band
and a hole at the Fermi surface.  At $T=0$, the low-energy
contributions to the spin-spin correlation function decay faster than
the high-energy contributions. In contrast, numerical studies seem to
suggest that at high temperatures the oscillating terms are still
present, but the slowest decaying term has pure power-law decay with
no oscillations \cite{FabriciusMcCoy,SirkerDiff}. This slowly decaying
term has been interpreted as due to spin diffusion at high
temperatures.  Here we will show that a diffusive term is already
present at low temperatures. In the following we use the boson
propagator in Eq.~(\ref {SelfE}), ~(\ref{Pi2ndorder})  to calculate the $q\sim0$ low-energy
contribution to the autocorrelation function in the regime $T\ll J$.

We can write the low-energy, long-wavelength contribution to $\bra
S_{l+x}^z(t)S^z_l(0)\ket$ as\be 
G(x,t)\equiv
-2\int_{-\infty}^{+\infty}\frac{d\omega}{2\pi}\int_{-\infty}^{+\infty}\frac{dq}{2\pi}\,e^{i(qx-\omega
  t)}\frac{{\rm Im}\chi_{\rm ret}(q,\omega)}{1-e^{-\omega/T}}, \label{Gxt}\ee
with $\chi_{\rm ret}(q,\omega)$ given by Eq.~(\ref {SelfE}) and (\ref{Pi2ndorder}) .  Note that the $-b\omega^2$ and $cv^2q^2$ terms in the 
self-energy simply rescale the energy and wave-vector by $(1+b)$ and $(1+c)$ respectively, giving a $T$-dependent 
velocity, $\tilde v\equiv \sqrt{(1+c)/(1+b)}v$. Doing the
integral over $q$ first, we find (see appendix \ref{appSpin} for details)
\begin{widetext}
\bea
\!\!\!\!\!\!\!\! \!\!\!\!\! G(x,t)&=&\frac{K}{2\pi v^2}(1+b)^{-\frac{1}{2}}(1+c)^{-\frac{3}{2}}\int_{-\infty}^{+\infty}\frac{d\tilde{\omega}}{4\pi} \,\exp[-i\tilde{\omega}\tilde{t}-i(\tilde{\omega}^2+2i\tilde{\gamma}\tilde{\omega})^{1/2}|\tilde{x}|/v]\,\frac{(\tilde{\omega}^2+2i\tilde{\gamma}\tilde{\omega})^{1/2}}{1-e^{-\tilde{\omega}/\tilde{T}}}-(\tilde{\gamma}\to-\tilde{\gamma}),
\label{Gxtintegral}
\eea
\end{widetext}
where $\tilde{t}=t(1+b)^{-1/2}$, $\tilde{x}=x(1+c)^{-1/2} $,
$\tilde{\gamma}=\gamma(1+b)^{-1/2}$ and $\tilde{T}=T (1+b)^{1/2}$. The
integral in Eq.~(\ref{Gxtintegral}) has both pole and branch cut
contributions, inside the light cone when $|t|>\tilde v|x|$, but only
the pole contribution outside the light cone when $|t|<\tilde v|x|$.
While this integral could be evaluated more generally numerically, we
focus on a couple of simple regions where we can obtain analytic
results. One of these is $|x|/v\lll 1/\gamma \ll t$ where \be
G(x,t)=(1+b)^{-1/2}(1+c)^{-3/2}[G_{0}(x,t)+G_{\rm int}(x,t)], \ee with
\bea
\label{LL_term}
G_0(x,t)&=&\frac{K}{8\pi^2v^2} \l\{\frac{\pi\tilde{T}}{\sinh[\pi\tilde{T}(\tilde{t}-\tilde{x}/v)]}\r\}^2 \\
&+&\frac{K}{8\pi^2v^2}
\l\{\frac{\pi\tilde{T}}{\sinh[\pi\tilde{T}(\tilde{t}+\tilde{x}/v)]}\r\}^2
\nn \eea is the Luttinger liquid result in terms of the rescaled
variables and\be 
\label{Spin_diff_term}
G_{\rm int}(x,t)=\frac{K\tilde{T}}{ v^2}\sqrt{\frac{\gamma}{2\pi
    t}}\,e^{-\gamma\tilde{x}^2/2v^2\tilde{t}} \ee is the diffusive
term.  The other simple region is $|t|<|x|/v \ll 1/\gamma$ where we
just obtain the Luttinger liquid result, $G_0$.  Note that even for
the static correlation function there are for $|x|/v\gtrsim 1/\gamma$
in principle $\gamma$-dependent corrections to the approximately
exponential decay (see appendix \ref{appSpin}). However, since
$1/\gamma\gg \xi$ where $\xi=v/(2\pi T)$ is the correlation length,
these corrections occur in a regime where the correlation function is
already extremely small and where also higher order corrections to
(\ref{LL_term}) have to be taken into account. A schematic
representation of the behavior expected in the different regions is
shown in Fig.~\ref{Diffusion}.
\begin{figure}[t!]
\includegraphics*[width=1.0\columnwidth]{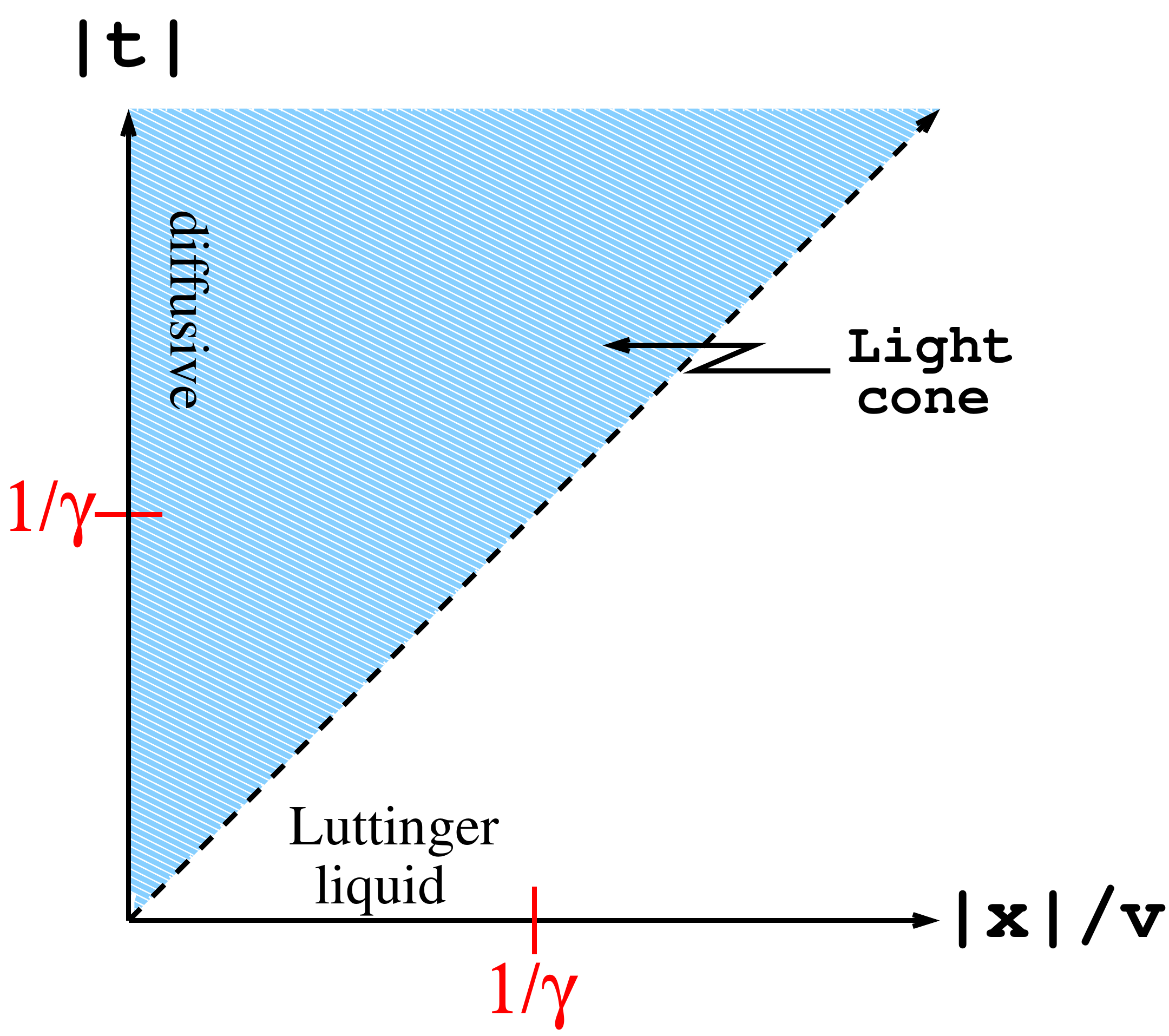}
\caption{The low-energy, long-wavelength contributions to $\langle
  S^z_{l+x}(t)S^z(0)\rangle$ at finite temperatures in different
  regions of the spacetime diagram. Inside the light cone,
  Eq.~(\ref{Gxtintegral}) has both branch cut and pole contributions
  but only pole contributions outside the light cone. The Luttinger
  liquid result is valid for $|t|<|x|/v \ll 1/\gamma$ whereas
  the diffusive term (\ref{Spin_diff_term}) dominates for $|x|/v\lll
  1/\gamma \ll t$.}
\label{Diffusion}
\end{figure}
We expect these results to be universally valid at large $x$ and $t$,
ignoring a possible Drude weight term, since the Fourier transform is
dominated by the small $\omega$ and $q$ region, at large $t$ and $x$.
$G_{\rm int}$ has the classic diffusion form.  Note that diffusion
occurs in a more limited domain than originally proposed for high-$T$
ferromagnets\cite{Bloembergen,deGennes,SteinerVillain} however.  We
have shown it to occur at low $T$, where $\gamma \ll T$, ignoring a
possible non-zero Drude weight term in the self-energy, but only
inside the light cone at $|x|/v\ll 1/\gamma \ll |t|$ and only for the
uniform part of the spin correlation function. (There is also a
power-law oscillating term.)  Importantly, the domain where classical
diffusive behavior occurs does include the self-correlation function.
For all $0<\Delta\leq1$, at finite temperatures and sufficiently long
times the autocorrelation function becomes dominated by a
\emph{low-energy} term\be \bra S_{l}^z(t)S^z_l(0)\ket \sim
T\sqrt{\frac{\gamma(T)}{t}},\quad (T\neq0, t\gg 1/\gamma) \ee with the
universal power-law decay $t^{-1/2}$ expected for diffusion in one
dimension.\cite{Bloembergen,deGennes,SteinerVillain} Note that the
more limited diffusive behavior we have found is related to the
Lorentz invariance of the underlying low energy effective Lagrangian.
The $b$ and $c$ terms break Lorentz invariance but only produce an
unimportant$T$-dependent shift of the velocity.  The important
breaking of Lorentz invariance, which leads to diffusion, is due to
the finite temperature. While diffusive behavior occurs only in a
limited space-time domain, it is sufficient to give diffusive behavior
in certain NMR experiments, as we show in the next sub-section.

\subsection{Spin-lattice relaxation rate}
\label{appD}
If the $q\sim 0$ contribution dominates the dynamics, diffusive
behavior should show up as characteristic frequency and magnetic field
dependence in NMR experiments. Before deriving a prediction for the
spin-lattice relaxation rate based on the results from the previous
section and comparing with experiment, we first want to point out a
general relation useful to take the finite magnetic field in NMR
experiments into account properly.

The linear response formula for the spin-lattice relaxation rate is\cite{Moriya56}
\be
\frac{1}{T_1}=\frac{1}{2}\int\frac{dq}{2\pi}\,|A(q)|^2S^{+-}(q,\omega_N)|_h,
\ee where $A(q)$ is the hyperfine coupling form factor, $\omega_N$ is
the \emph{nuclear} magnetic resonance frequency and \be
S^{+-}(q,\omega)|_h=\frac{1}{N}\sum_{l,l^\prime}\int_{-\infty}^{+\infty}dt\,
e^{i\omega_Nt}\bra S_l^+(t)S_{l^\prime}^-(0)\ket|_h
\label{transverseS} \ee is the transverse dynamical spin structure
factor. Here $S^\pm_l=S^x_l\pm iS^y_l$ are the raising and lowering
spin operators. The expression in Eq.~(\ref{transverseS}) is to be
calculated using Hamiltonian (\ref{HXXZspin}) in the presence of a
magnetic field $h$. We focus on the experimentally relevant Heisenberg
point $\Delta=1$.  We would like to express $1/T_1$ in terms of the
longitudinal structure factor \be
S^{zz}(q,\omega)=\frac{1}{N}\sum_{l,l^\prime}\int_{-\infty}^{+\infty}dt\,
e^{i\omega_Nt}\bra S_l^z(t)S_{l^\prime}^z(0)\ket \ee at zero field.
The latter is more easily calculated in the field theory since $S^z_l$
is related to the local density of fermions $n_l$, whereas $S_l^{\pm}$
have nonlocal representations in terms of Jordan-Wigner fermions.
Although the exchange term in the Heisenberg model is isotropic, the
magnetic field term in Eq.~(\ref{HXXZspin}) breaks rotational
symmetry. As a result, $S^{+-}(q,\omega)|_h$ cannot be directly
replaced by $2S^{zz}(q,\omega)|_h$ at finite field. However, we note
that the longitudinal field has a trivial effect on $S_l^{\pm}(t)$\be
S^{\pm}_l(t)=e^{iHt}S_l^{\pm}e^{-iHt},=e^{-i\omega_et}e^{i\bar{H}t}S_l^{\pm}e^{-i\bar{H}t},
\ee where $\bar{H}=H(h=0)$ and $\omega_e=\mu_Bh$ is the
\emph{electron} magnetic resonance frequency. If we assume in addition
that $T\gg \omega_e$, the magnetic field dependence in the thermal
average can be neglected and we have \be S^{+-}(q,\omega)|_h\approx
\frac{2}{N}\sum_{l,l^\prime}\int_{-\infty}^{+\infty}dt\,
e^{i(\omega_N-\omega_e)t}\bra S_l^z(t)S_{l^\prime}^z(0)\ket|_0, \ee
where the correlation function is calculated at $h=0$. This leads to
the expression for the spin-lattice relaxation rate\be
\frac{1}{T_1}\approx\int\frac{dq}{2\pi}\,|A(q)|^2S^{zz}(q,\omega_N-\omega_e)|_0.
\ee Using $\omega_e\gg\omega_N$ and \be
S^{zz}(q,-\omega_e)=\frac{2\,\textrm{Im }\chi_{\rm
    ret}(q,\omega_e)}{1-e^{\omega_e/T}}, \ee we find in the regime
$\omega_e\ll T$\be
\frac{1}{T_1}\approx-\frac{2T}{\omega_e}\int\frac{dq}{2\pi}\,|A(q)|^2\textrm{Im
}\chi_{\rm ret}(q,\omega_e).\label{1T1formula} \ee

If the integral in Eq.~(\ref{1T1formula}) is dominated at low
temperatures, $T\ll J$, by the $q\sim 0$ mode of $\chi_{\rm
  ret}(q,\omega)$ then we can perform the momentum integral using the
retarded correlation function in Eq.~(\ref{SelfE}). Assuming
furthermore that the momentum dependence can be neglected, $A(q\sim
0)\equiv A=\mbox{const}$, and using the appropriate parameters
(\ref{selfE_parameters_iso}) for the isotropic case we find \be
\label{T1_6}
\frac{1}{T_1T} =  2|A|^2 X_1 \sqrt{\frac{X_2}{2}+\sqrt{\l(\frac{X_2}{2}\r)^2+\l(\frac{\gamma}{\omega_e}\r)^2}}  ,
\ee
where \bea
X_1&=&1+\frac{g}{2}-\frac{g^2}{8}+\frac{5g^3}{64}+\frac{3\sqrt{3}}{2\pi}T^2,\nn\\
X_2&=&1+\frac{g^2}{4}-\frac{g^3}{32}\l(3-\frac{8\pi^2}{3}\r)+\frac{\sqrt{3}}{\pi}T^2.
\eea
In the limit $\gamma(T)\gg \omega_e$, we obtain the diffusive behavior \be
\frac{1}{T_1T}\sim \sqrt{\frac{\gamma(T)}{\omega_e}}\sim \sqrt{\frac{T/\ln^2(J/T)}{\omega_e}}.
\label{1T1short}
\ee

In the NMR experiment by Thurber {\it et al.},
Ref.~[\onlinecite{ThurberHunt}], the copper-oxygen spin chain compound
Sr$_2$CuO$_3$ was studied. In this compound an in-chain oxygen site
exists such that $A(q)=A\cos(q/2)$ in (\ref{1T1formula}). In NMR
measurements on this oxygen site any contribution to the spin-lattice
relaxation rate coming from $q\sim\pi$ is therefore almost completely
suppressed so that the $q\sim 0$ mode dominates. Note that for NMR
measurements on the copper or the apical oxygen site both low-energy
contributions would be present with the $q\sim\pi$ being dominant. In
Ref.~[\onlinecite{ThurberHunt}] the experimental data for the in-chain
oxygen site have been interpreted in terms of a spin-lattice
relaxation rate $1/T_1T\sim T/\sqrt{\omega_e}$. While the frequency
dependence does agree with that found in Eq.~(\ref{1T1short}) and
generally expected if diffusion holds, we note that the temperature
dependence of the effective diffusion constant $D_s\equiv v^2/2\gamma$
is different.

To quantitatively compare our prediction with experiment we note that
the form factor $A(q)=A\cos(q/2)$ for the in-chain oxygen site is
given by \be 
\label{hyperfine}
|A|^2=\frac{k_B}{2\hbar}\frac{(2C^b)^2+(2C^c)^2}{\pi^3
  k_B^2J^2}(g\gamma_N\hbar)^2, \ee where $k_B$ is the Boltzmann
constant, $C^{b,c}$ are the dimensionless components of the hyperfine
coupling tensor, $g\gamma_N\hbar = 4.74\times 10^{-9}$ eV and $J$ is
the exchange coupling measured in Kelvin. The components of the
hyperfine coupling tensor can be obtained by performing a $K-\chi$
analysis, i.e., a comparison between measurements of the Knight shift
$K$ and the magnetic susceptibility $\chi$. Such an analysis has been
performed in Ref.~[\onlinecite{ThurberHunt}] leading to $2C^b=95 \pm
10$ and $2C^c=44 \pm 10$ kOe/$\mu_B$. Furthermore, the exchange
coupling $J$ has to be determined. We note that a parameter-free field
theory formula for the static magnetic susceptibility follows from
\bea
\label{mag_sus}
\chi &=& -\chi_{\rm ret}(q,0)=\frac{K}{2\pi v}\frac{1}{1+c} \\
&\approx &
\frac{1}{\pi^2}\l[1+\frac{g}{2}+\frac{3g^3}{32}+\frac{\sqrt{3}}{\pi}
T^2\r] \nn \eea where we have used Eqs.~(\ref{SelfE},
\ref{Pi2ndorder}). In the second line, we have specialized for the
isotropic point, with $v=\pi/2$, $K\to 1+g/2+g^2/4+g^3/8$, and the
parameter $c$ as given in (\ref{selfE_parameters_iso}).
Eq.~(\ref{mag_sus}) is in agreement with the result in
Refs.~[\onlinecite{Lukyanov,SirkerLaflorencie2}]. Using this formula
and fitting the susceptibility data in
Ref.~[\onlinecite{MotoyamaEisaki}] we obtain $J=2000 \pm 200$ K. In
Fig.~\ref{Fig_T1} the shaded area denotes the region covered by the
curves obtained by varying the parameters $J$ and $C^{b,c}$
independently within the given range. However, if we believe that the
field theory describes the zero temperature limit correctly then the
possible variation of the theoretical results is strongly
overestimated. From Eq.~(\ref{T1_6}) we see that $1/(T_1T)\to 2|A|^2$
for $T\to 0$ and is therefore not affected by the question how large
$\gamma$ is. For a given $J$ we can therefore fix $(2C^b)^2+(2C^c)^2$
in (\ref{hyperfine}) by fitting the experimental data at low
temperatures. The remaining variation is then very small (see the two
examples in Fig.~\ref{Fig_T1}).
\begin{figure}[t!]
\includegraphics*[width=1.0\columnwidth]{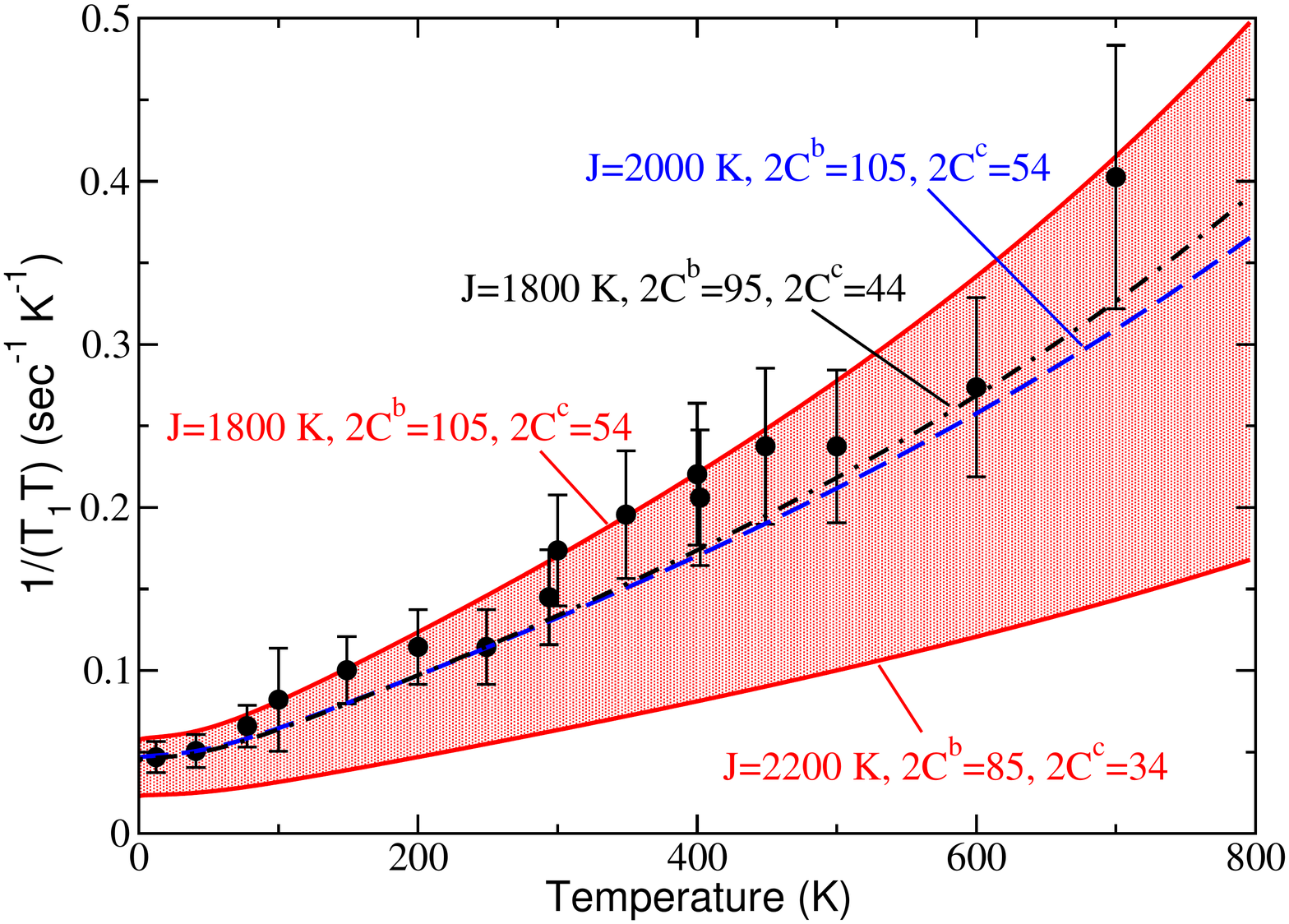}
\caption{Experimental data for the spin-lattice relaxation rate of the
  spin chain compound Sr$_2$CuO$_3$ at $h=9$ T taken from
  Ref.~[\onlinecite{ThurberHunt}] (dots) compared to our theory. The
  solid lines represent the two extreme limits when varying $J$,
  $C^{b,c}$ independently within the given error bars. The other two
  curves correspond to fixing the hyperfine constants for a given $J$
  by the experimental data at low temperatures.}
\label{Fig_T1}
\end{figure}
For $\gamma=0$ corresponding to purely ballistic transport, as
suggested by the Bethe ansatz calculation of
Ref.~[\onlinecite{KluemperJPSJ}], $1/(T_1 T)$ would be approximately
constant.\cite{SirkerPereira} Further support that the good agreement
seen in Fig.~\ref{Fig_T1} is not accidental is obtained by an analysis
of the NMR data for the apical oxygen site which is also presented in
Ref.~[\onlinecite{ThurberHunt}]. In this case the hyperfine coupling
tensor $A(q)$ is momentum independent and the contribution from
low-energy excitations with $q\sim\pi$ dominates. In this case field
theory yields\cite{SachdevNMR} \be
\label{T1_staggered}
\frac{1}{T_1}=\frac{(D^b)^2+(D^c)^2}{(2\pi)^{3/2}}\frac{(g\gamma_N\hbar)^2}{\hbar
  k_B J} \ee with $D^b=23\pm 10$ and $D^b=14\pm 10$
kOe/$\mu_B$.\cite{ThurberHunt} Using again $J=2000$ K the experimental
data in Ref.~[\onlinecite{ThurberHunt}] are very well described by
this formula. Therefore the experimental data for the magnetic
susceptibility and the NMR relaxation rates, testing the low-energy
contributions of $\chi_{\rm ret}(q,\omega)$, {\it both} at $q\sim 0$ and
$q\sim \pi$, are consistently described using $J\sim 2000$ K and a
hyperfine coupling tensor as determined experimentally by a $K-\chi$
analysis.

\subsection{Consequences for electron spin resonance}
\label{ESR}
Here we want to point out a connection between our study of the
conductivity of the $S=1/2$ Heisenberg chain and a previous theory of
electron spin resonance (ESR) for such systems.\cite{OshikawaAffleck}
By also using a self-energy approach for the boson propagator it was
shown in this work that the ESR lineshape is Lorentzian with a width
given by the imaginary part of the retarded self-energy. In
particular, the case of an exchange anisotropy perpendicular to the
applied magnetic field was considered. In this case the ESR linewidth
for an applied magnetic field $h$ is given by
\begin{equation}
\label{linewidth}
\eta =-\mbox{Im}\, \Pi_u^{\rm ret}(h,h)/(2h)
\end{equation}
where $\Pi_u^{\textrm{ret}}(h,h)$ is the retarded self-energy due to Umklapp
scattering, Eq.~(\ref{SelfE2}), for the isotropic case. From
(\ref{Pi2ndorder}) and (\ref{selfE_parameters_iso}) we therefore
immediately obtain
\begin{equation}
\label{linewidth2}
\eta \equiv \gamma =\frac{\pi}{2}g^2T .
\end{equation}
This is consistent with Eq.~(6.27) in
Ref.~[\onlinecite{OshikawaAffleck}] using the standard replacement
$\lambda \to g/4$ in reverse and in addition $\lambda \to v\lambda$
with $v=\pi/2$ because in [\onlinecite{OshikawaAffleck}] the spin
velocity was set to $1$.  The RG flow for the running coupling
constant $g$ is now cut off by the larger of the temperature $T$ or
the applied magnetic field $h$.  If these two scales are sufficiently
different, we can use Eq.~(\ref{coup_g}) for $g$ with $T$ being
replaced by $\mbox{max}(T,h)$. In this case (\ref{linewidth2}) is a
parameter-free prediction for the ESR linewidth of a $S=1/2$
Heisenberg chain with a small exchange anisotropy perpendicular to the
applied magnetic field.

We want to stress that a vanishing relaxation rate $\gamma$ would
dramatically affect the ESR linewidth. From (\ref{SelfE2}) and
(\ref{F_func4}) we obtain for $\omega/T,\, q/T\ll 1$ \bea
\label{ESR_expansion}
\Pi_u^{\rm ret}(q,\omega) &\sim & F^{\rm ret}(q,\omega)-F^{\rm ret}(0,0) \\
&\sim& \omega^2+q^2 +\frac{q^2\omega^2}{T^2} + \im \l(2\gamma\omega  +\frac{q^2\omega}{T}\r) \nn .
\eea
This leads to 
\begin{equation}
\label{ESR_expansion2}
\eta =-\mbox{Im}\, \Pi_u^{\rm ret}(h,h)/(2h) \sim \gamma + h^2/T 
\end{equation}
and therefore a linewidth $\eta\sim h^2/T$ if $\gamma=0$.

\subsection{Consequences for the spin structure factor}
\label{Sqw}
The longitudinal dynamical spin structure factor at finite
temperatures can be obtained from 
\begin{equation}
\label{Sqw1}
S^{zz}(q,\omega)= -\frac{2}{1-\e^{-\beta\omega}}\mbox{Im}\, \chi^{\textrm{ret}}(q,\omega) 
\end{equation}
In the low-energy, long-wavelength limit the retarded spin-spin
correlation function can be again expressed by the boson propagator
using Eq.~(\ref{SelfE}). With the help of (\ref{Kubo2}) we therefore
obtain a direct relation between the spin structure factor and the
real part of the conductivity
\begin{equation}
\label{Sqw2}
S^{zz}(q,\omega)= \frac{2q^2}{\omega(1-\e^{-\beta\omega})}\sigma^\prime(q,\omega) \quad (q\ll 1). 
\end{equation}
>From (\ref{fullSigma}) we see that our theory predicts a Lorentzian
lineshape at finite temperatures. For $T\to 0$ the parameters
$b,c,\gamma$ in (\ref{fullSigma}) vanish and we find the well-known
free boson result $2S^{zz}(q,\omega)=K|q|\delta(\omega-v|q|)$.

Here some comments are in order. In our perturbative calculation we
have only included the band curvature terms to first order. While the
lowest order contribution vanishes at zero temperature, the terms of
order $2n$ show divergencies on shell, $\omega\sim
vq$.\cite{PereiraSirkerJSTAT} While it is not yet clear how to sum up
a series of such terms, we know from BA\cite{BougourziKarbach} that
they lead to a finite linewidth $\delta\omega(q)\sim q^3$ at zero
temperature. The lineshape at zero temperature is distinctly
non-Lorentzian with threshold singularities.  For finite temperature
such that $\gamma(T)\gg \delta\omega(q)$ we can, however, neglect
these terms and Eq.~(\ref{Sqw2}) becomes valid. Furthermore, we note
that the second order Umklapp contribution to the self-energy leads
for zero temperature to the high-frequency tail of $S^{zz}(q,\omega)$
given in Eq.~(7.24) of Ref.~[\onlinecite{PereiraSirkerJSTAT}]. From
Eq.~(\ref{Sqw2}) it is also clear that the regular part of the
conductivity and a part which could possibly yield a ballistic channel
for dc transport are not independent but rather have to fulfill sum
rules, as for example, $\int d\omega
S^{zz}(q,\omega)=2\pi\sum_j\exp(\im q j)\langle S_j(0)S_0(0)\rangle $.
The right hand side of this sum rule can be calculated
numerically with high accuracy since it only involves static correlation
functions.

\section{Haldane-Shastry model}
\label{secHS}
Unlike the $XXZ$ model, generic spin models with exchange interactions beyond nearest
neighbor are not integrable. Here we want to consider a special spin model with
long-range interactions that is known to be integrable, even though
it is not solvable by Bethe ansatz. The Haldane-Shastry model is given
by\begin{equation}
H=\sum_{j<l}J_{jl}\mathbf{S}_{j}\cdot\mathbf{S}_{l},\label{eq:HSmodel}\end{equation}
where $J_{jl}=J(j-l)=\{(L/\pi)\sin[\pi(j-l)/L]\}^{-2}$
is the long-range exchange interaction.\cite{HaldaneHSmodel1988, ShastryPRL1988} 
For a chain with size $L$ and periodic boundary conditions,
$J_{jl}$ is inversely proportional to the square of the chord distance
between sites $j$ and $l$. We set the energy scale of the exchange
interaction such that in the thermodynamic limit $J_{jl}\to(j-l)^{-2}$. 

The Haldane-Shastry model is completely integrable as the transfer
matrix satisfies the Yang-Baxter equation. The conserved quantities are nonlocal and are not obtained by the traditional method of expanding  the
transfer matrix. However, they  have been found using the ``freezing trick''
starting from the SU(2) Calogero-Sutherland model.\cite{Talstra1995} The exact spectrum is given by the
dispersion of free spinons, which contrasts with the interacting spinons
of the Heisenberg model. The dynamical structure factor has been calculated
exactly.\cite{Talstra1994} It has a square-root singularity at the lower threshold of
the two-spinon continuum, as expected for SU(2) symmetric models.
Contrary to the Heisenberg model, there are no multiplicative logarithmic
corrections to the asymptotic behavior of correlation functions. This
means that in the effective field theory for the Haldane-Shastry model
the coupling constant of the umklapp operator is fine tuned to zero.

Since the umklapp operator is responsible for the leading decay
process of the spin current in the calculation in Sec. \ref{appF}, we
may expect that spin transport in the Haldane-Shastry model is purely
ballistic.  This seems reasonable given the picture of an ideal spinon
gas, but requires that the coupling constants of all higher order
Umklapp processes also vanish exactly.

Although the total spin is conserved, the spin density does not satisfy
a local continuity equation due to the long-range nature of the exchange
interactions. Nonetheless, the spin current operator can be defined
using the equivalence to a model of spinless fermions  coupled to an electromagnetic field. In terms of Jordan-Wigner
fermions, the Haldane-Shastry model reads\bea
H&=&\sum_{j<l}J_{jl}\left[\left(n_{j}-\frac{1}{2}\right)\left(n_{l}-\frac{1}{2}\right)\right.\nonumber\\
&&\left.+\frac{(-1)^{l-j}}{2}\left(c_{j}^{\dagger}e^{i\pi\sum_{p=j}^{l-1}n_{p}}c^{\phantom\dagger}_{l}+h.c.\right)\right].\eea
We then consider a magnetic flux $\Phi$ that threads the chain
and couples to the fermionic fields in the form $c_{j}^{\dagger}c^{\phantom\dagger}_{l}\to c_{j}^{\dagger}c^{\phantom\dagger}_{l}e^{-i\sum_{m=j}^{l-1}A_{m,m+1}}$.
Here $A_{m,m+1}$ is the vector potential defined on the link between
sites $m$ and $m+1$, such that $\sum_{m=1}^{N}A_{m,m+1}=\Phi$.
The Hamiltonian in the presence of the electromagnetic field becomes
\bea
H&=&\sum_{j<l}J_{jl}\left[\left(n_{j}-\frac{1}{2}\right)\left(n_{l}-\frac{1}{2}\right)\right.+\frac{(-1)^{l-j}}{2}\times\nonumber\\
&&\left.\times\left(c_{j}^{\dagger}e^{i\sum_{p=j}^{l-1}(\pi n_{p}-A_{p,p+1})}c^{\phantom\dagger}_{l}+h.c.\right)\right].\eea
The current operator associated with a given link is defined as\bea
\mathcal{J}_{j,j+1} & = & -\frac{\partial H}{\partial A_{j,j+1}}\nonumber\\
 & = & \frac i2\sum_{n>0}\sum_{\ell=0}^{n-1}(-1)^{n}J(n)\: \times \\
 &&\times c_{j-\ell}^{\dagger}e^{i\sum_{p=j-\ell}^{j-\ell+n-1}(\pi n_{p}-A_{p,p+1})}c^{\phantom\dagger}_{j-\ell+n}\nonumber\\
 &&+h.c.. \nonumber \eea
Therefore the integrated current operator in the limit of weak fields,
$A_{j,j+1}\to0$, reads\begin{eqnarray}
\mathcal{J} & = & \sum_{j}\mathcal{J}_{j,j+1} \\
 & = & \frac i2\sum_{j}\sum_{n\neq0}(-1)^{n}nJ(n)c_{j}^{\dagger}e^{i\pi\sum_{p=j}^{j+n-1}n_{p}}c^{\phantom\dagger}_{j+n}.\nonumber 
\label{eq:currentJHS}\end{eqnarray}
Reverting to spin operators, we obtain the spin current operator\begin{equation}
\mathcal{J}=\frac{i}{2}\sum_{j}\sum_{n\neq0}nJ(n)S_{j}^{+}S_{j+n}^{-}.\end{equation}
Notice that $\mathcal{J}$ is a nonlocal operator, since it acts on
sites separated by arbitrary distances. 

We can now compute the commutator of the current operator with the
Hamiltonian in Eq. (\ref{eq:HSmodel}). The result is\bea
[\mathcal{J},H]&=&\frac{i}{2}\sum_{l\neq m\neq n}[(m-n)J_{mn}(J_{ln}-J_{lm})\nonumber\\
&&+(2l-m-n)J_{lm}J_{ln}]S_{l}^{z}S_{m}^{+}S_{n}^{-}.\eea
In the thermodynamic limit, $J(n)\to1/n^{2}$ and we find\begin{equation}
[\mathcal{J},H]=0.\end{equation}
Therefore, the current operator is conserved and transport is ballistic in the thermodynamic limit. That the conservation law does not hold for a finite chain can be easily seen by considering a three site ring. In this case the Hamiltonian and current operator of the Haldane-Shastry model reduce to those of the Heisenberg model, and these two operators clearly do not commute.

The current operator is orthogonal to the set of conserved quantities associated with integrability. For instance, the first nontrivial conserved quantity is\begin{equation}
Q_{3}=i\sum_{ijk}\frac{z_{i}z_{j}z_{k}}{z_{ij}z_{jk}z_{ki}}\mathbf{S}_{i}\cdot(\mathbf{S}_{j}\times \mathbf{S}_{k}),\end{equation}
where $z_{j}=e^{i2\pi j/L}$ and $z_{ij}=z_{i}-z_{j}$.  The orthogonality  can be shown by arguing that the current operator is odd under a $\pi$ rotation about the $x$ axis, which takes $S^z_j\to -S^z_j$ and $S^y_j\to -S^y_j$, whereas the conserved quantities  derived in Ref. [ \onlinecite{Talstra1995}]  are SU(2) invariants and therefore even under this transformation. Unlike the current operator, these conserved quantities commute with the Hamiltonian for a finite chain with periodic boundary conditions.

\section{The attractive Hubbard model}
\label{Charge}
We now want to consider the negative $U$ Hubbard model at 1/2-filling
\be H=\sum_j[-(c^\dagger_{\alpha j}c_{\alpha j+1}+h.c.)-U(\hat n_j-1/2)^2] .\ee

We may take the continuum limit and bosonize the resulting Dirac
fermions, which now carry a spin index, finally introducing charge and
spin bosons.  These are decoupled up to irrelevant operators.  For
$U<0$ and zero magnetic field, the spin excitations are gapped and we
expect the associated contribution to the low frequency conductivity
to be Boltzmann suppressed at temperatures small compared to the spin
gap.  The low energy effective Hamiltonian contains only the charge
boson, which we again call $\phi$ for simplicity. This low-energy
Hamiltonian now contains the Umklapp scattering term \bea H_u&=&\tilde
\lambda \int dx\left[
  \psi^\dagger_{L\uparrow}\psi^\dagger_{L\downarrow}
  \psi_{R\uparrow}\psi_{R\downarrow} + h.c.\right]\nonumber\\
&=&\lambda \int dx \cos (\sqrt{8\pi}\phi ).\eea The bare coupling
constant $\lambda$ is $\propto -U$ at small $U$. It is marginally
irrelevant, as in the spinless model at $\Delta =1$, and the effective
coupling at scale $T$ behaves similarly to Eq.~(\ref{coup_g}) \be
{1\over \lambda (T)}+{\ln \lambda (T)\over 2}=\ln \left[ {T_0\over
    T}\right]\ee where the cut-off scale $T_0$ depends on $U$. Thus
the self-energy for the charge boson has the form of
Eq.~(\ref{Pi2ndorder}), with \be 2\gamma(T) = \pi \lambda^2(T)T\ee as
in Eq.~(\ref{selfE_parameters_iso}).

This lowest order, RG improved, calculation again ignores the
possibility of a Drude weight.  A rigorous Mazur bound on the Drude
weight for the Hubbard model was established in
[\onlinecite{ZotosPrelovsek}], using the conserved energy current, but
it fails at half-filling where $\langle\mathcal{J}\mathcal{J}_E\rangle=0$.  In
this case, other local conserved charges have not been considered.

In the limit $U/t\to -\infty$, the spin gap becomes infinite and the
model is equivalent to the Heisenberg model with $J\propto t^2/U$ and
charge operators replacing spin operators. In this limit, we may
import all results discussed above for the Heisenberg model. The fact
that the Hubbard interaction is marginal ($K=1$), as well as the
equivalence with the Heisenberg model at large $U$ follow from the
exact duality transformation
\bea c_{\uparrow j}&\to& (-1)^jc^\dagger_{\uparrow j}\nonumber \\
c_{\downarrow j}&\to&c_{\downarrow j} \eea which changes the sign of
$U$.  This maps bilinear operators as follows
\bea \hat n^j&\to& 2S^z_j\nonumber \\
\hat c^\dagger_{j,\alpha}c_{j+1,\alpha}-h.c.&\to &
-(c^\dagger_{j}\sigma^zc_{j+1}-h.c.).  \eea The charge current maps
into the spin current. Adding a chemical potential to dope away from
1/2-filling at negative $U$ is equivalent to adding a magnetic field
at positive $U$.  In the large negative $U$ limit, where we only allow
empty sites and double occupancy, this becomes equivalent to the
Heisenberg model ($\Delta =1$). The spin SU(2) symmetry of the $U>0$
model implies a ``hidden'' SU(2) symmetry in the charge sector. Thus,
the Hubbard model at half-filling actually has $SU(2)\times SU(2)/Z_2
=SO(4)$ symmetry for either sign of $U$.

The continuum limit model also has a duality symmetry
\bea \psi_{L/R\uparrow}(x)&\to& \psi_{L/R \uparrow}^\dagger (x)\nonumber \\
\psi_{L/R\downarrow}(x)&\to& \psi_{L/R \downarrow}(x).
\eea
Umklapp at $U<0$ maps into the $\pm$ component of the marginal spin operator at $U>0$:
\be \psi^\dagger_{L\downarrow}\psi^\dagger_{L\uparrow}\psi_{R\uparrow}\psi_{R\downarrow}
\to \psi^\dagger_{L\downarrow}\psi_{L\uparrow}\psi^\dagger_{R\uparrow}\psi_{R\downarrow}.\ee
Similarly, the product of left and right charge currents 
maps into the product of left and right z-components of spin currents: 
$J_LJ_R\to 4J^z_LJ^z_R$.  There is an exact $SU(2)\times SU(2)$ symmetry of 
 the continuum model. Umklapp and $J_LJ_R$ interactions 
are sitting exactly on the separatrix of the Kosterlitz-Thouless RG flow.

We might think of a negative $U$ as arising from phonon exchange. More
general models, with longer range interactions will not be on the
separatrix but may have the low energy Hamiltonian, with Umklapp and
$J_LJ_R$ interactions. For a range of interaction parameters, the
Umklapp will be irrelevant and we get a Lorentzian conductivity.

The negative $U$ Hubbard model may be thought of as being in a
one-dimensional version of a superconducting state. However, it is
important to realize that it is not a true superconductor and the
presence or absence of a finite Drude weight at finite $T$ is
independent of the absence of true superconductivity. One way of
seeing this point is to consider a ring of circumference $L$
penetrated by a flux $\Phi$. If the ring was a torus of sufficiently large 
thickness, we might be able to model its properties using London
theory. Using the London equation \be \vec j = -{n_s(e^*)^2\over
  mc}\vec A\ee and $A=\Phi /L$, we obtain a persistent current
$\propto 1/L$ \be j=-{n_s(e^*)^2\Phi\over mcL}.\label{London}\ee
Here $e^*=2e$ is the charge of the Cooper pairs.  In London theory,
this formula is true at finite $T$ with $n_s(T)$, the superfluid
density, a decreasing function of $T$ which is non-zero for $T<T_c$.
On the other hand, for the one-dimensional Hubbard model, the finite
size spectrum for current-carrying states is: \be E={\pi v_c\over
  L}(n+\Phi /2\pi )^2+E_0\ee where $v_c$ is the velocity of charge
excitations. (This result is obtained ignoring the irrelevant Umklapp
interactions.)  The partition function is \be
Z=Z_0\sum_{n=-\infty}^\infty \exp [-(\pi v_c/LT)(n+\Phi /2\pi )^2],\ee
and the resulting persistent current \be j=\partial F/\partial \Phi
.\ee For sufficiently large $L$ such that $T$ is much greater than the
finite size gap, $T\gg v_c/L$, this gives \be j\approx -2\sin \Phi
\sqrt{LT\over v_c}e^{-\pi LT/v_c}.\ee This is exponentially small in
$L$, unlike the case for a superconductor, Eq. (\ref{London}). The
superfluid density vanishes at any finite $T$.  On the other hand, if
we ignore Umklapp scattering, the model has a $T$-independent Drude
weight $D=v_c/4\pi$, as in Eq.~(\ref{D0}).

>From the London equation for a superconductor it follows that \be
\lim_{\omega \to 0}\omega \ \hbox{Im} \ \sigma (\omega
,q)={n_s(e^*)^2\over m}\ee independent of $q$. On the other hand, for
the Hubbard model, $\mbox{Im}\,\sigma(q,\omega)$ can be obtained from
Eq.~(\ref{opticalcond}) with $v\to v_c$ and it is easy to see that
$\omega\,\mbox{Im}\,\sigma(q,\omega)$ vanishes in the limit $\omega\to
0$ at non-zero $q$.

\section{Summary and conclusions}
\label{Conclusions}
To summarize, we have studied how nontrivial conservation laws affect
the transport properties of one-dimensional quantum systems. We
focused, in particular, on the spin current in the integrable $XXZ$
model. Away from half-filling, a part of the spin current is protected
by conservation laws leading to a ballistic channel which coexists
with a diffusive channel. For the half-filled case, however, none of
the infinitely many local conservation laws responsible for
integrability has any overlap with the current by symmetry.
Nonetheless, several works have argued in favor of ballistic transport
at finite temperatures even in this situation. This would require the
existence of an unknown nonlocal conservation law which has finite
overlap with the current operator.  To investigate this controversial
problem, our strategy was the following: Assuming that such a
conservation law does not exist, we calculated the current relaxation
within the Kubo formalism using a self-energy approach for the boson
propagator. Since the parameters in the field theory are explicitly
known due to the integrability of the microscopic model we obtained a
parameter-free formula for the optical conductivity. Any shift of
weight from this regular into a Drude part can then be described
within the memory-matrix formalism and explicitly tested for by
comparing the parameter-free result with numerical and experimental
data.

By a numerical study of the time-dependent current-current correlation
function we have shown that the intermediate time decay is well
described by the calculated relaxation rate. This excludes, in
particular, the large Drude weight found by Bethe ansatz in
Ref.~[\onlinecite{KluemperJPSJ}]. As we have shown, this result
corresponds to our field theory with the relaxation rate set exactly
to zero by hand. A small Drude weight can, however, not be excluded by
the intermediate time data. This includes, in particular, the Drude
weight found in a different Bethe ansatz calculation.\cite{Zotos} At
least at high temperatures it is, however, known that this approach
violates exact relations so that the solution cannot be exact. It is
not clear at present if or why these results could be considered as an
approximate solution at low temperatures. Quantum Monte Carlo
calculations, on the other hand, are performed using imaginary times.
The results presented in Refs.~[\onlinecite{AlvarezGros,AlvarezGros2}]
therefore cannot be used to resolve a decay rate smaller than the
separation of Matsubara frequencies. An interesting possibility is the
idea to analytically continue our results to imaginary frequencies and
to check if this is consistent with Quantum Monte Carlo data. Such an
analysis has been performed recently for the isotropic case and good
agreement was found.\cite{GrossjohannBrenig} It would be very
interesting to perform a similar analysis in the anisotropic case for
$\Delta$ values such that the decay rate is not too small. The
advantage would be that additional complications due to logarithmic
corrections present at the isotropic point do not occur. In
Ref.~[\onlinecite{SirkerPereira}] we also presented numerical data for
the current-current correlation funciton at infinite temperatures and
showed that even in this case a large time scale persists. It is
therefore unclear how data obtained by exact diagonalization can be
reliably extrapolated to the thermodynamic limit.

>From our perturbative result for the boson propagator at finite
temperatures we found that the long-wavelength contribution to the
spin-lattice relaxation rate is diffusive. By comparing with
experiments on the spin chain compound Sr$_2$CuO$_3$ we showed that
our formula describes the experimental data well. Importantly, we
showed that the data for the magnetic susceptibility and the $q\sim 0$
and $q\sim \pi$ contributions to the spin-lattice relaxation rate are
all consistently described by the field theory formulas with the same
exchange constant $J$ and using the hyperfine coupling tensor as
determined experimentally.  We also pointed out that our results are
consistent with a previous theory of electron spin resonance in spin
chains\cite{OshikawaAffleck}. For the longitudinal spin structure
factor at zero magnetic field our theory predicts a crossover from the
known non-Lorentzian lineshape with linewidth $\sim q^3$ at zero
temperature, to a Lorentzian lineshape with a linewidth set by the
relaxation rate $\gamma$ at sufficiently high temperatures.

We also discussed the integrable Haldane-Shastry model where the
spectrum is known to consist of free spinons in contrast to the
interacting spinons in the Heisenberg model. As in the Heisenberg
model at zero magnetic field the current operator does not have any
overlap with the known conserved quantities. However, in this case the
nonlocal current operator itself becomes conserved in the
thermodynamic limit and transport is therefore ballistic.

Finally, we showed that our calculations for the spin current in the
$XXZ$ model carry over to the charge current in the attractive Hubbard
model at half-filling. We stressed the point that an infinite dc
conductivity does not imply that the system is a true superconductor.
This can be seen by calculating the superfluid density which turns out
to be zero in this case.

\appendix
\section{Spin Green's function}
\label{appSpin}
Consider a boson propagator with a momentum-independent decay term (see Eq.~(\ref{SelfE}))
\bea \chi_{ret}(q,\omega)&\equiv& -i\int_0^\infty dt\int_{-\infty}^\infty dx \nonumber \\
 &\times& e^{i(\omega t-qx)} \langle [\partial_x\phi (t,x),\partial_x\phi (0,0)]\rangle_T  \nonumber \\
&\to &{q^2\over \omega^2-q^2+2i\gamma (T)\omega},
\label{Gret}\eea
for small $\omega$ and $q$. (We set the velocity equal to 1.  $\gamma$
must be $>0$ by causality.)  We wish to calculate the long-time
behavior of the correlation function \be G(x,t)\equiv \frac{K}{2\pi}
\langle\partial_x\phi (t,x)\partial_x\phi (0,0)\rangle_T ,\ee which
can be expressed in terms of $\chi_{ret}(q,\omega)$ with the help of
Eq.~(\ref{Gxt}). Using this equation and (\ref{Gret}) we find
\begin{widetext}
  \be G(x,t)\to -i\int{d\omega dq\over (2\pi )^2}{e^{-i(\omega
      t-qx)}q^2\over (1-e^{-\beta \omega})[q-(\omega^2+2i\gamma \omega
    )^{1/2}][q+(\omega^2+2i\gamma \omega )^{1/2}]} -(\gamma \to
  -\gamma ).\label{Cw}\ee We now consider only the first term; we
  return to the $\gamma \to -\gamma$ term later. We are defining
  $(\omega^2+2i\gamma \omega )^{1/2}$ to be a particular branch of the
  square root; let us specify carefully which one. It will be
  convenient to define this branch with the branch cut along the
  negative imaginary $\omega$ axis between $\omega =0$ and $\omega
  =-2i\gamma$. Thus \bea (\omega^2+2i\gamma \omega )^{1/2}&\to&
  \sqrt{2\gamma u-u^2} \ \ (\omega =-iu+\delta ,\ \ 0<u<2\gamma ,\ \
  \delta \to 0^+)
  \nonumber \\
  &\to & \omega +i\gamma \ \  (|\omega|\gg \gamma )\nonumber \\
  &\to& -\sqrt{2\gamma u-u^2} \ \ (\omega =-iu-\delta ,\ \ 0<u<2\gamma
  ,\ \ \delta \to 0^+).  \eea
\end{widetext}
We do the $q$-integral first.  Noting that the result is manifestly an
even function of $x$, we just consider explicitly the case $x>0$.
Noting that this branch of $(\omega^2+2i\gamma \omega)^{1/2}$ always
has a positive imaginary part for real $\omega$, we may close the $q$
integral in the upper half plane, encircling the pole at
$q=(\omega^2+2i\gamma \omega )^{1/2}$, giving
\bea G_{\hbox{first}}(x,t) &\to& \int {d\omega \over 4\pi }\exp [-i\omega t+i(\omega^2+2i\gamma \omega )^{1/2}|x|] \nonumber \\
&\times & {(\omega^2+2i\gamma \omega )^{1/2}\over 1-e^{-\beta
    \omega}}.\label{C}\eea Consider the analytic
structure of the integrand in the complex $\omega$ plane. (We consider
only the first term; we return to the $\gamma \to -\gamma$ term
later.)  There is a branch cut along the negative imaginary axis from
$\omega =0$ to $\omega =-2i\gamma$.  There are also poles at $\omega
=2\pi inT$ for $n=\pm 1$, $\pm 2, \ldots$. The integrand behaves as
$1/\omega^{1/2}$ at $\omega \to 0$ so the integral is convergent.
Let's assume $|t|>|x|$. Then, we can close the integral in the upper
half plane for $t<0$ or the lower half-plane for $t>0$. First consider
the simpler case $t<0$.  Then we only pick up the contributions from
the poles $\omega =2\pi inT$, $n=1$, $2$, $3, \ldots$ Let us now
assume $\gamma (T)\ll T$. Note that we expect this to be true at low
$T$, $\gamma (T)\propto T^{4K-3}$, $K>1$. Then, at the poles: \be
(\omega^2+2i\gamma \omega )^{1/2}\approx \omega = 2\pi inT.\ee Then,
for $t<0$, the first term gives the approximately $\gamma$-independent
result: \bea
 G_{\hbox{first}}^{\, t<0}(x,t)\!\!\! &\to&\!\!\! -\pi T^2\sum_{n=1}^\infty
n\, e^{2\pi nT (t-\sqrt{1+\gamma /(\pi nT)}|x|)} \\
&\approx & { -e^{-\gamma |x|}\over 4\pi}{1\over \{(1/\pi T)\sinh [\pi
  T(t-|x|)]\}^2}. \nonumber
\eea Actually, while we can always ignore
the $\gamma$-dependent corrections to the pre-factor, those
corrections in the exponent cannot be ignored at sufficiently large
$|x|$.  We have Taylor expanded the square root in the exponential to
first order in $\gamma /T$.  We see that the correction becomes
important at $|x|$ of order $1/\gamma$. At such large values of $|x|$,
higher order corrections must also be included.  If we assume
$|x|\ll 1/\gamma$ we may simply drop the $e^{-\gamma |x|}$ factor.

Now consider the first term in Eq.~(\ref{C}) for $t>0$.  There are now
both pole and cut contributions.  For the pole contributions we find
again \be G_{\hbox{first,pole}}^{\, t>0}\to {-e^{-\gamma |x|}\over
  4\pi}{1\over \{(1/\pi T)\sinh [\pi T(t-|x|)]\}^2}.\ee But now, we
must also consider the cut contribution. We may Taylor expand the
denominator in Eq.~(\ref{C}) to first order since $\beta |\omega |\ll
1$ along the cut, $\omega =-iu$, with $0<u<2\gamma \ll T$. Thus: \bea
G_{\hbox{first,cut}}^{\, t>0} &\to& {1\over 4\pi \beta}\int_0^{2\gamma }{du\over
  u}\sqrt{2\gamma u-u^2} e^{-ut}\\
 &\!\!\!\!\!\!\!\!\!\!\!\!\!\!\!\!\!\!\! \times &\!\!\!\!\!\!\!\!\!\!\!\!\!\! \{ \exp [-i\sqrt{2\gamma
  u-u^2}|x|]+\exp [i\sqrt{2\gamma u-u^2}|x|]\}. \nonumber
\label{cut}\eea Now let us assume $|t|\gg 1/\gamma$, so that the
integral is dominated by $u\ll 1/\gamma$ and we may approximate it by:
\bea G_{\hbox{first,cut}}^{\, t>0} &\to& {1\over 2\pi \beta}\int_0^{\infty
}du\sqrt{2\gamma \over u} e^{-ut} \\
&\times & \cos \l[\l(\sqrt{2\gamma
  u}-\frac{u^{3/2}}{2\sqrt{2\gamma}}\r)|x|\r].\nonumber \eea Here we have
expanded the square root to first order in the exponential. This first
correction, as well as the higher order ones, may be dropped provided
that: \be |x|\ll |t|\sqrt{\gamma |t|}.\label{xcond}\ee Note that this
is automatically true since we have already assumed $|x|<|t|$ and
$\sqrt{\gamma |t|}\gg 1$.  Assuming Eq. (\ref{xcond}) and letting
$u=v^2/2$, this becomes \bea C_{\hbox{first,cut}}^{\, t>0} &\to& 
{\sqrt{\gamma}\over 2\pi \beta}\int_0^{\infty }dve^{-v^2|t|/2}[
e^{-i\sqrt{\gamma}|x|v}+c.c.] \nonumber \\
&=& T\sqrt{2\pi \gamma \over |t|}e^{-\gamma
  x^2/(2|t|)}. \eea

Now consider the $\gamma \to -\gamma$ term in Eq.~(\ref{Cw}). We
define a convenient branch of $\sqrt{\omega^2-2i\gamma \omega}$ with
the branch cut along the positive imaginary axis from $\omega =0$ to
$\omega =2i\gamma$ obeying:
\begin{widetext}
\bea (\omega^2-2i\gamma \omega )^{1/2}&\to& \sqrt{2\gamma u-u^2} \ \  (\omega =iu+\delta ,\ \ 
0<u<2\gamma ,\ \  \delta \to 0^+)
\nonumber \\
&\to & \omega -i\gamma \ \  (|\omega|\gg \gamma )\nonumber \\
&\to& -\sqrt{2\gamma u-u^2} \ \  (\omega =iu-\delta ,\ \ 
0<u<2\gamma ,\ \  \delta \to 0^+)
\eea
\end{widetext}
Now, choosing again $x>0$, the $q$ integral, in the upper half plane, encircles 
the pole at $q=-(\omega^2-2i\gamma \omega )^{1/2}$, giving:
\bea G_{\hbox{second}}(x,t) &\to& \int {d\omega \over 4\pi }\exp [-i\omega t-i(\omega^2-2i\gamma \omega )^{1/2}|x|] \nonumber \\
&\times & {(\omega^2-2i\gamma \omega )^{1/2}\over 1-e^{-\beta
    \omega}}.\label{C2second}\eea
Now the pole terms give
\be G_{\hbox{second,pole}}\to 
{-e^{\gamma |x|}\over 4\pi}{1\over \{(1/\pi T)\sinh [\pi T(t+|x|)]\}^2}\ee
(for either sign of $t$). The cut term, is now present for $t<0$, giving:
\be
G_{\hbox{second,cut}}^{\, t<0} = G_{\hbox{first,cut}}^{\, t>0} \to T\sqrt{2\pi \gamma \over |t|}e^{-\gamma x^2/(2|t|)}
\ee
Combining both terms gives
\be G(x,t)\to G_0(x,t)+G_{int}(x,t) \label{0int},\  \  (|t|>|x|).\ee
where $G_0(x,t)$ is the result in the non-interacting case
\bea G_0(x,t)\to &-& {1\over 4\pi } \left\{{e^{-\gamma |x|}\over \{(1/\pi T)\sinh [\pi T(t-x)]\}^2}\r. \nonumber \\
&+&\l. {e^{\gamma |x|}\over \{(1/\pi T)\sinh [\pi T(t+x)]\}^2}\right\}\eea
and
\be G_{int}(t,x)\to T\sqrt{2\pi \gamma \over |t|}e^{-\gamma x^2/(2|t|)}\label{Cint}\ee
(for both signs of $t$). 

Note that we assumed $|t|>|x|$ below Eq.~(\ref{C}) determining in which 
half-plane the contour for the $\omega$-integral was closed.  We may attempt to 
evaluate the integrals using the same method also when $|x|>|t|$. Now the 
$\omega$ integral is closed in the upper half plane for the first term and 
the lower half plane for the second, regardless of the sign of $t$.  Thus the 
cut terms do not appear and we only get the pole terms:
\be G(x,t)\to G_0(x,t)\  \  (|t|<|x|).\label{Cxgt}\ee

Let's reiterate the various conditions on $t$ and $x$ in order for
Eqs.~(\ref{0int}) and (\ref{Cxgt}) to hold.  The form of $G_0$ depends
on the assumption $|x|\ll 1/\gamma$ but not an any particular
assumption on $t$. On the other hand, the form of $G_{int}$, which is
only present for $|t|>|x|$, also assumes $|t|\gg 1/\gamma$.

\begin{acknowledgments} 
  The authors thank A.~Alvarez and C.~Gros for sending us their
  quantum Monte Carlo data and acknowledge valuable discussions with
  T.~Imai, A.~Kl\"umper, A.~Rosch and X.~Zotos. This research was
  supported by NSERC (I.A.), CIfAR (I.A.), the NSF under Grant No.
  PHY05-51164 (R.G.P.), and the MATCOR school of excellence (J.S.).
\end{acknowledgments}


\end{document}